**Vapor-liquid-solid growth of unconventional nanowires**

Thang Pham* and Arindom Nag
Department of Materials Science and Engineering, Virginia Tech, Blacksburg, VA 24060
* Corresponding author: thangpham@vt.edu

**Abstract**

Vapor-liquid-solid (VLS) growth has delivered a level of synthetic control in group IV and III-V semiconductor nanowires that remains unmatched in any other class of one-dimensional materials. For Si, Ge, and GaAs, decades of mechanistic study have made diameter, growth direction, crystal phase, and axial composition into design parameters. For oxides, carbides, and chalcogenides, including emerging van der Waals materials, which we refer to collectively as non-conventional nanowires, one-dimensional structures have been reported in abundance, but comparable deterministic control has not been achieved despite their predicted functional properties and broad application potential. This review examines where and why the two groups diverge. We survey and categorize the literature on VLS and VLS-related synthesis of non-conventional nanowires, following the sequence of the growth process itself, namely precursor creation and delivery, seed particle formation and evolution, and nucleation and growth of one-dimensional structures. Each step is compared against the group IV and III-V baseline to locate shared behavior and genuine differences. From this comparison we identify seed particles, spanning the choice of seed material, the chemical composition of the seed during and after growth, and its dynamics, as the principal bottleneck restricting deterministic synthesis, with precursor creation and delivery acting as an upstream constraint that narrows the accessible seed design space. We discuss potential solutions, including new precursor chemistries, improved reactor designs, and *in situ* characterization of growth, that would deepen mechanistic understanding of nanowires already demonstrated by VLS and extend seeded growth to emerging quantum and functional materials, ultimately paving the way for deterministic synthesis and integration of complex one-dimensional nanomaterials.

## Introduction

Nanowires of group IV (Si, Ge)[1,2] and III-V (GaAs, InAs)[3,4] semiconductors can now be synthesized with a degree of control that approaches design. Extensive *ex situ* and *in situ* efforts have produced detailed understanding of their growth mechanisms[5,6], enabling selection of growth direction[7], crystal phase[8,9], and axially modulated composition, including heterostructures and superlattices with abrupt interfaces [6,10,11]. This capability rests on the vapor-liquid-solid (VLS) mechanism, and it has motivated wide exploration of other material classes, including oxides[12–16], carbides[17–24], and chalcogenides[25–32], with the goal of extending both the mechanistic framework and the application space of one-dimensional materials[33–38]. The outcome of that exploration is uneven. One-dimensional structures of these materials have been reported in large numbers, yet control over morphology, crystal phase, and composition remains far from what the conventional systems achieve. Understanding the origin of this gap, and identifying the step of the growth process where it arises, is the purpose of this review.

The VLS framework was proposed by Wagner and Ellis to explain Au-seeded growth of Si whiskers[39] and consists, in its canonical form, of three steps[5,40,41] (Fig. 1a): delivery of a vapor-phase precursor to a substrate bearing metal particles, formation of a liquid alloy droplet governed by the eutectic composition and temperature, and supersaturation-driven nucleation of the solid at the droplet interface. The binary phase diagram of the seed and nanowire materials rationalizes each step (Fig. 1b). Subsequent work, particularly *in situ* transmission electron microscopy (TEM)[42,43], (Fig. 1c) confirmed the three steps in nucleation and growth of seeded nanowires. Additionally, it also revealed important deviations from the simple 3-step picture, including growth well below the nominal eutectic temperature[44] and nucleation behavior that depends sensitively on droplet geometry and gas environment[8]. These observations, discussed in Section 3.1.2, converted the qualitative framework into quantitative mechanism, and that conversion is what made deterministic synthesis possible in the conventional systems.

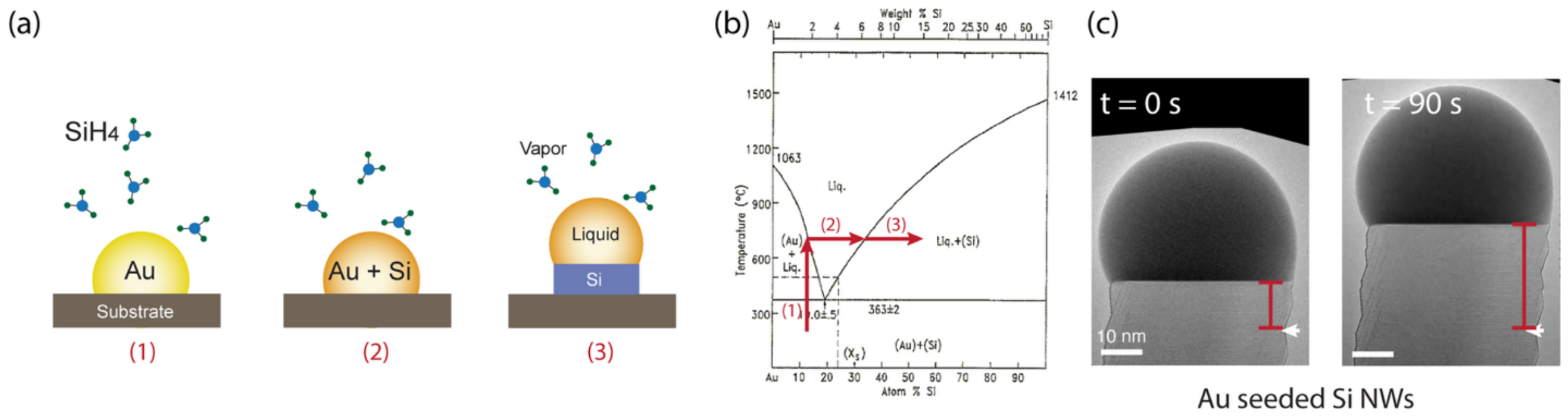

**Figure 1. Vapor-liquid-solid (VLS) growth of one-dimensional nanomaterials**. (a) Main steps of the growth process: (i) Vaporization of sources (solid, gas or liquid precursors) and its transportation to the substrate. (ii) Precursor delivery to the seed particle site, including both atomic and molecular delivery. (iii) Nucleation and continued growth of a nanowire at the droplet-nanowire interface. (b) Binary phase diagram of Au-Si showing the three steps involving in a VLS synthesis. (c) *In situ* and atomic resolution TEM images of Au-seeded Si nanowire in ETEM at 500°C with $5\times10^{-5}$ Torr disilane[42]. The white arrows mark a reference point to show the growth of the same nanowire captured by *in situ* TEM. (c) Adapted with permission from ref[42]. Copyright 2016 Royal Society of Chemistry.

Many non-conventional nanowire systems do not satisfy the assumptions this framework takes for granted. Suitable volatile precursors are often unavailable, and low-temperature eutectics between the nanowire constituents and common seed materials are frequently inaccessible. A variety of modified and hybrid growth strategies have emerged that relax one or more of these constraints while retaining the essential elements of seeded growth. These developments motivate a comparative perspective, in which the conventional systems serve as a baseline for evaluating how differences in precursor chemistry, seed particle behavior, and interfacial kinetics change growth outcomes in more complex materials.

Our review aims to uncover general principles governing VLS synthesis across nanowires with different crystal structures, bonding characteristics, and chemical compositions. The discussion follows the order in which each step takes place during growth. Section 1 addresses precursors and their delivery, Section 2 the choice and evolution of seed particles, and Section 3 the resulting one-dimensional structures and their growth mechanisms. Each section begins with a concise account of the group IV and III-V baseline, examines shared and distinct behavior in oxides, carbides, and chalcogenides, and closes with challenges and opportunities. Section 4 extends the discussion to van der Waals nanowires, an emerging subset of the chalcogenides whose intrinsic structural anisotropy modifies what the seed particle must accomplish, and whose recent development illustrates both the reach and the present limits of seeded growth. Table 1 maps the material classes covered in this review together with the precursor routes, seed materials, and morphologies demonstrated for each, and its columns follow the organization of Sections 1 to 3. Section 3 closes with Table 2, which summarizes the contrast between conventional and non-conventional systems at every stage of the growth process.

Across these sections, a consistent picture emerges. The three stages of VLS growth are coupled rather than independent. Precursor chemistry and reactor design set which seed compositions and growth temperatures can be reached, and the seed particle in turn governs which structures, phases, and compositions can form. Within this coupling, we identify the seed particle, from the choice of seed material to its chemical composition during and after growth and its dynamics, as the principal bottleneck restricting deterministic synthesis of non-conventional nanowires. We support this assessment throughout the review and discuss potential solutions, including new precursor and source chemistries, reactor architectures with independently controlled precursor delivery, and *in situ* growth experiments adapted to these material classes. These directions serve two goals. The first is to resolve the growth mechanisms of the many nanowires that VLS has already produced but not yet explained. The second is to extend seeded growth to emerging quantum and functional materials, where the controls a seed particle provides, over growth location, dimension, and composition, would add capability to materials so far grown only by other routes.

**Table 1**. Material classes of non-conventional nanowires discussed in this review, with the precursor routes, seed materials, and one-dimensional morphologies demonstrated for each. Rows are ordered by composition and, within the chalcogenides, by the structural unit bounded by the van der Waals gap. For the final group, one-dimensional morphologies are established by non-VLS growth.

| **Materials group and representative systems** | **Precursors (Section 1)** | **Seed materials (Section 2)** | **Demonstrated 1D morphologies (Sections 3 and 4)** | **Refs.** |
|---|---|---|---|---|
| **Oxides** ZnO, $SnO_2$, SnO, $In_2O_3$, ITO, $Ga_2O_3$, MgO, MnO, | Carbothermic reduction [12,13,16,47]; | Au and other metals (Ag, Cu, Pt, Fe) | Wires [12-16,84]; belts and ribbons [51,66,102]; rings | 12-16, 45, 47-51, 60- |
| CaO, NiO, $Sm_2O_3$, $Eu_2O_3$, $Al_2O_3$, $GeO_2$, $MoO_3$, $WO_3$ | hydrogen-assisted vaporization [50]; combined atomic metal and $O_2$ [47]; PLD within flux window [15,45,101]; salt activation for Mo and W oxides [79,80]; electron-beam generation *in situ* [92,93] | [14,69,84]; Ge [73,75-77]; self-seeded Sn and Al [85,92]; Sn [93]; alkali salts [78-80] | and springs by VS [115,116]; twisted wires and tubes [74,110]; radial core-shell only, no axial heterojunctions [15,45,62,63] | 63, 66, 69, 72-80, 84, 85, 92-96, 101, 102, 110, 115, 116 |
| **Carbides** SiC, TiC, TaC, $B_4C$ | Carbothermic reduction [17,19,22]; combined atomic metal and molecular carbon source ($CH_4$, $CH_3SiCl_3$, orthocarborane) [18,20,59] | Fe, Ni, and other group VIII metals [23]; Au less common | Wires [18-20,23,24]; nanosprings and helical structures [21,22,59] | 17-24, 59 |
| **Monochalcogenides (layered vdW)** GaS, GaSe, GeS, GeSe, $GeSe_2$, $GeS_{1-x}Se_x$ | Direct thermal evaporation of chalcogenide powders; hydrogen-assisted vaporization [53-56,129] | Au, Ag [53-55]; Bi [132] | Wires with both layer orientations [53]; edge-grown ribbons with uniform stacking [131]; twisted and helical wires [56,111,112,135]; twin-plane bicrystals [131]; GeSe-$GeSe_2$ core-shell [132]; mixed-dimensional 1D-2D heterostructures [133]; alloys [112,134] | 52-56, 111, 112, 129, 131-135 |
| **Layered chalcogenides (multi-atom slabs)** $Bi_2Se_3$, $Bi_2Te_3$, $In_2Se_3$ | Thermal evaporation and hydrogen-assisted vaporization [29,30,57]; molecular precursors by MOCVD [25,27,28] | Au | Wires, ribbons, and plates, frequently coexisting [25–32,57]; phase transformation to ternary wires [57] | 25-32, 57 |
| **Non-layered chalcogenides** PbS, PbSe, ZnSe, CdSSe | Thermal evaporation from solid sources [90,108,109] | Au [90,108,109] | Eshelby-twisted wires and branched | 90, 103, |
| | | | hierarchical structures [108,109]; surface-guided planar wires [90]; alloyed wires [103] | 108, 109 |
| **Dichalcogenide monolayer ribbons** $MoS_2$, $MoSe_2$, $WS_2$ | Salt-generated volatile intermediates (sodium molybdate) [78,82] | Molten alkali salt droplets (NaCl, KCl, hydroxides) [78,82,83] | Monolayer and few-layer ribbons [82,83]; core-shell architectures [81] | 78, 81-83 |
| **Transition metal trichalcogenides** $NbS_3$, $NbSe_3$, $TiS_3$ ($TaS_3$, $ZrS_3$ as bulk) | Salt-generated volatile intermediates [58]; CVT for bulk crystals [130] | Multicomponent Au-alkali halide alloy droplets [58] | Wires and ribbons with distinct growth orientations, lateral dimension set by droplet size [58] | 58, 130 |
| **Emerging quasi-1D vdW and quantum materials, non-seeded to date** $Ta_2Pd_3Se_8$, $Ta_2Ni_3Se_8$, $Nb_2Se_9$, $Bi_4I_4$, $Sb_2S_3$, GaSeI, InSeI, SnIP, SnTe, $Cd_3As_2$, TaAs, WP, $BaTiS_3$, $BaZrS_3$ | CVT [140]; liquid-precursor CVD [141]; solid-state growth [136]; uncatalyzed vapor-phase crystallization [138,139] | None established; in $Bi_4I_4$, Au particles act as VS nucleation sites or, at specific compositions, as VLS seeds [138] | Wires, ribbons, and quasi-2D sheets selected by growth parameters or seed composition [138,139, 146, 149, 150]; helical and twisted morphologies [137] | 136-142, 146, 149, 150 |

## 1. Precursors

### 1.1 Background

In this section, we discuss the first step in VLS growth: the creation of precursor vapor and its transport to seed particle sites on the growth substrate. There are two general routes to create and supply precursors to the substrate: (i) atomic precursors generated by physical methods and (ii) molecular precursors delivered by chemical methods (Fig. 2).

In the first route, precursors are typically inorganic compounds, such as elemental metals or metal oxides, that exhibit low vapor pressures under ambient conditions. These precursors are physically atomized either through thermal activation (thermal evaporation[12,13,16], often above 500°C and in many cases well above 1000°C) or by high-energy irradiation, for example pulsed laser deposition[15,45]. To reduce the thermal budget and increase precursor volatility, a reducing agent is frequently mixed with the solid precursor[13]. The physical route can be further divided into sub-categories, discussed below, based on the specific approach used to generate high-pressure vapor from solid sources (Figs. 2a-b). The vaporized species are transported to the substrate by a carrier gas and incorporated into the growing nanowire at the seed particle-nanowire (i.e., liquid-solid or L-S) interface.

In the chemical route, precursors are typically metal-organic compounds (Fig. 2c) that exist as gases, liquids, or solids at ambient conditions but exhibit high vapor pressures at relatively low temperatures[21,28]. These molecular species are delivered to and dissociate on multiple surfaces[46], but preferentially on the seed particle surface (the vapor-liquid or V-L interface) due to its enhanced chemical activity, generating the atomic constituents required for nanowire growth. In general, molecular precursors offer greater flexibility for creating complex structures, particularly heterostructures and superlattices composed of dissimilar materials. Their relatively broad temperature windows for dissociation enable sequential growth of multiple components with clean and abrupt interfaces through gas switching using mass flow controllers and fast on/off shutters. However, the chemical route also presents notable challenges, including the limited availability of suitable metal-organic precursors and the complexity of the required synthesis infrastructure. These systems often require toxic or highly combustible gas handling as well as high-vacuum reactors. In contrast, physical methods benefit from simpler experimental setups and the wide availability of solid, inorganic precursors.

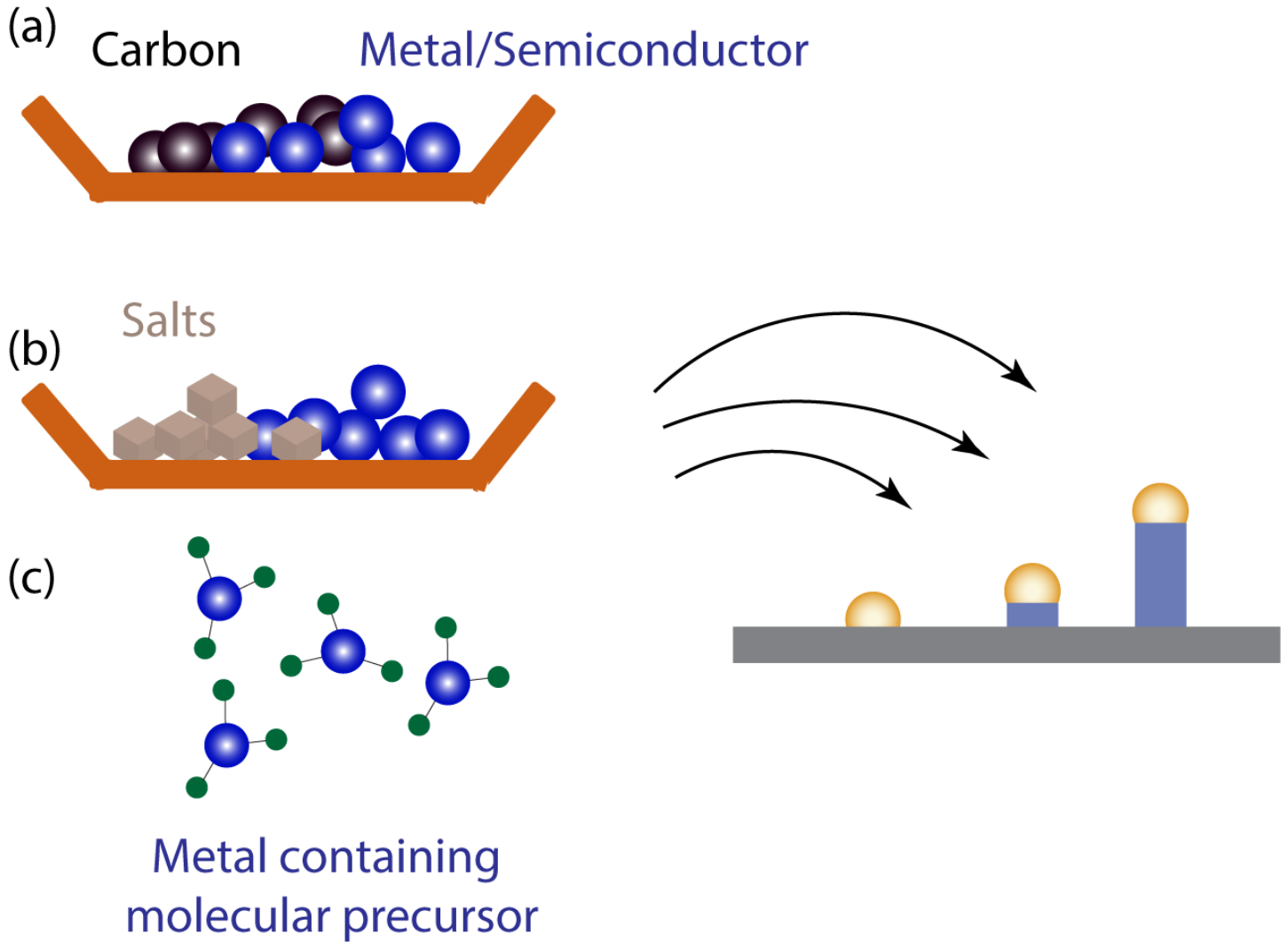

**Figure 2.** Precursors in VLS synthesis. (a-b) Physical methods of supplying atomic precursors: (a) carbothermic reduction of metals (semiconductors) and their oxides. (b) Salt-assisted metal precursor evaporation. (c) Molecular precursors.

For conventional nanowires, as we defined here as group IV and III-V systems, such as Si, Ge, GaAs, InAs, GaN, both precursor delivery routes have been extensively explored, with particular emphasis on molecular precursors due to their versatility. Common molecular precursors include silane ($SiH_4$) and disilane ($Si_2H_6$) for Si, germane ($GeH_4$) and digermane ($Ge_2H_6$) for Ge, trimethylgallium ($Ga(CH_3)_3$) for Ga, and arsine ($AsH_3$) for As. These precursors have been successfully employed to grow nanowires with corresponding chemical compositions, such as Si, Ge and GaAs nanowires, even axial heterostructures and superlattices of dissimilar materials, such as Si-Ge[10,11] and GaAs-InAs[6]. In contrast, for oxides, carbides and chalcogenides, as we refer as non-conventional nanowires in this review, there are comparatively few examples employing molecular precursors. Instead, most non-conventional nanowires have been synthesized using physical approaches in which solid inorganic precursors are thermally activated and transported to the growth substrate. In the following sections, we detail the major categories within this route.

### 1.2 Physical methods of supplying atomic precursors

In this subsection, we introduce several physical approaches used to create and supply atomic precursors, including carbon-assisted vaporization, hydrogen-assisted vaporization, salt-assisted vaporization, and hybrid methods that combine physical and chemical pathways. Each approach offers distinct advantages and limitations. Table 1 summarizes which routes have been applied to each class.

***Carbon-assisted vaporization***

Carbothermic reduction is a widely used method for metal vapor delivery[12,13,16], in which carbon reduces metal oxides to produce metal vapor at high temperatures (typically 800-1000°C) (Fig. 2a). Various carbon sources have been employed, including carbon black, activated carbon, and carbon nanotube powders, all of which can act as effective reducers[16]. The method is widely used in the synthesis of oxide nanowires with ZnO nanowires representing the prototypical and most extensively studied example. In a typical synthesis, ZnO powder is mixed with carbon powder and heated to high temperature, where carbothermic reduction occurs:

$$C_{(s)} + ZnO_{(s)} \rightarrow CO_{(g)} + Zn_{(v)} \qquad (1)$$
$$Zn_{(v)} + 1/2\ O_{2(g)} \rightarrow ZnO_{(nanowire)} \qquad (2)$$

The resulting Zn vapor is transported by a carrier gas (Ar or $N_2$) to the substrate and incorporated into the seed particle droplet (for example, Au). It was suggested that oxidation of Zn at the droplet-substrate (L-S) interface leads to crystallization of ZnO layers and subsequent nanowire growth. Some recent studies suggest that the oxygen source for this oxidation step does not necessarily originate from CO or $CO_2$ produced during reaction (1), but rather from residual $O_2$ and/or $H_2O$ in the growth environment, either through leaky connections[47] or due to limited vacuum quality during growth at low pressure ($\sim 10^{-3}$ Torr) or ambient pressure. This straightforward approach has enabled the synthesis of a broad range of nanowires in addition to ZnO, including oxides ($SnO_2$[47], $In_2O_3$[48,49], $Ga_2O_3$ [50,51]), carbides ($B_4C$[17], SiC[22], TaC[19]), and chalcogenides ($Ga_2S_3$[52]).

***Hydrogen-assisted vaporization***

Hydrogen-assisted vaporization represents another physical route, in which hydrogen gas serves as the reducing agent instead of carbon. Compared to carbothermic reduction, this method generally operates at lower temperatures (400-800°C) but is applicable to a more limited range of materials. Examples include metal oxides such as $Ga_2O_3$[50], but mostly metal chalcogenides such as GaS[53], GaSe[54], GeS[55,56], $In_2Se_3$[57], and $Bi_2Se_3$[29]. Representative reactions for chalcogenide and oxide nanowire growth using hydrogen are:

$$H_{2(g)} + GeS_{(s)} \rightarrow Ge_{(v)} + H_2S_{(g)} \qquad (3)$$
$$Ge_{(v)} + H_2S_{(g)} \rightarrow GeS_{(nanowire)} + H_{2(g)} \qquad (4)$$

***Salt-assisted precursor activation***

In addition to carbon- and hydrogen-assisted routes, alkali metal halides have emerged as effective agents for activating transition-metal precursors in the VLS growth of non-conventional nanowires (Fig. 2b). When combined with elemental metals or metal oxides, salts such as NaCl promote the formation of volatile metal-containing intermediates, thereby increasing the effective vapor-phase flux of high-melting transition metals at moderate temperatures (600-800°C). This mechanism relaxes a central limitation of physical precursor delivery in oxides, carbides, and chalcogenides, namely the insufficient vapor pressure of many transition metals under typical growth conditions. Typical chemical reactions in salt-assisted growth are[58]:

$$2Nb_2O_{5(s)} + 3NaCl_{(s)} \rightarrow NbCl_3O_{(v)} + 3NaNbO_{3(l)} \quad (5)$$
$$NbCl_3O_{(v)} + 3S_{(g)} + 5/2H_2 \rightarrow NbS_{3(nanowire)} + 3HCl_{(g)} + H_2O_{(g)} \quad (6)$$

Importantly, in salt-assisted VLS the role of the salt is not limited to precursor volatilization. Post-growth chemical analysis in several systems reveals incorporation of alkali and halide species into the catalyst particle together with the metal seed and nanowire constituents, indicating that salt participation can extend to the catalyst droplet itself[58]. As a result, precursor activation and catalyst chemistry become intrinsically coupled. We note that this distinction is critical for mechanistic interpretation, as it differentiates salt-assisted VLS from salt-assisted vapor-solid growth schemes in which the salt serves only to enhance vapor transport without participating in catalyst-mediated growth.

From a practical perspective, salt-assisted precursor activation provides a modular pathway to access VLS growth regimes for material systems that do not readily form low-temperature eutectic composition with common seed metals (e.g., Au, Ag, Al) or lack suitable molecular precursors. As such, it complements carbon- and hydrogen-assisted vaporization while enabling access to new regions of growth parameter space for non-conventional nanowires.

***Combined atomic and molecular precursors***

A fourth approach combines atomic and molecular precursor delivery. In this scheme, atomic species are generated by thermal evaporation of low-melting-point metals, while molecular precursors such as oxygen ($O_2$) or hydrocarbons ($C_xH_y$) are introduced directly at the growth substrate, where they react at the seed particle to form oxide or carbide nanowires. The process can be described by the following reactions, in which M indicates transition metals:

$$M_{(s)} \rightarrow M_{(v)} \quad (7)$$
$$M_{(v)} + x/2O_{2(g)} \rightarrow MO_{x(nanowire)} \quad (8)$$
$$M_{(v)} + C_xH_y \rightarrow MC_{x(nanowire)} + y/2H_{2(g)} \quad (9)$$

To prevent premature reactions, the growth reactor typically employs separate gas lines that introduce oxygen or hydrocarbon precursors only near the substrate. Nanowires synthesized using this approach include $SnO_2$[47] (using Sn powder and $O_2$ gas), SiC[18] (using Si powder and methyltrichlorosilane, $CH_3SiCl_3$, gas), TiC[20] (using Ti powder and methane gas), and $B_4C$[59] (using B powder and orthocarborane gas).

## 1.3 Challenges and opportunities

Even though several non-conventional nanowires have been synthesized, the degree of control over their structure, composition, and spatial placement remains far below what has been achieved for group IV and III-V counterparts. Here, we propose two major differences in precursor creation and delivery between these two material classes that likely underlie the gap in deterministic synthesis.

The first barrier is the lack of suitable molecular precursors for the synthesis of oxides, carbides, and chalcogenides. In general, an effective molecular precursor for nanowire growth must satisfy two requirements:

(i) an appropriate dissociation temperature for VLS growth, and

(ii) compatibility with high-vacuum environments.

For nanowires in group IV (Si, Ge) and III-V (GaAs, InAs) systems, molecular precursor selection benefits directly from decades of research in thin-film deposition of the same materials. Indeed, molecular sources such as silane ($SiH_4$) for Si, germane ($GeH_4$) for Ge, trimethylgallium ($Ga(CH_3)_3$) for Ga, and arsine ($AsH_3$) for As, which have long been used in semiconductor thin-film growth, meet both requirements and function effectively for the conventional nanowire synthesis. In contrast, the pool of suitable molecular precursors for oxide, carbide, and chalcogenide nanowires is considerably more limited. Some transition-metal molecular precursors, such as diethylzinc (DEZ), which are commonly used in atomic layer deposition (ALD), may exhibit markedly different dissociation behavior, surface adsorption, and reaction chemistry at the elevated temperatures (>500°C) required for VLS growth compared to typical ALD conditions (from room temperature to below 100°C). This disparity complicates direct transfer of molecular precursor chemistries from thin-film deposition to VLS nanowire growth in the non-conventional material systems. In addition, some molecular precursors required for chalcogenide nanowire growth, such as sulfur- or selenium-containing gases, for example $H_2S$ or $H_2Se$, are detrimental to vacuum environments. These species readily react with reactor components and are difficult to remove by pumping, leading to degraded vacuum levels and severe cross-contamination. Collectively, these factors strongly restrict the use of molecular precursors for non-conventional nanowire growth.

The second challenge lies in the simplicity of the synthesis reactors commonly employed for the synthesis of non-conventional nanowires, namely horizontal tube furnaces. Although this type of reactor has enabled the growth of a wide range of oxide, carbide, and chalcogenide nanowires, it suffers from several inherent limitations. First, a typical tube furnace does not provide independent temperature control for different precursors or between precursors *vs*. the growth substrate. In a representative experiment, solid precursor powders (for example ZnO and carbon) and substrates (for example $SiO_2$/Si chips) are loaded into the same tube, and their temperatures are determined solely by their positions within the furnace, that is, by the intrinsic temperature gradient of the tube. This configuration poses significant challenges when materials exhibit distinct vapor pressures within a narrow temperature window: one precursor may evaporate excessively while another fails to vaporize at all. As a result, growth may be suppressed entirely, or the resulting nanowires may exhibit non-uniform stoichiometry and morphology due to poorly controlled atomic precursor flux.

Additionally, tube-furnace systems offer virtually no means to dynamically modulate fluxes of atomic precursors during growth, such as through the use of shutters. Consequently, it is difficult to progress from single-component nanowires to axial heterostructures by altering precursor supply. Indeed, only a few examples of axial heterostructured nanowires synthesized using atomic precursors have been reported for oxide nanowire systems[45,60–63]. This limitation calls for improved engineering solutions that integrate multiple, spatially separated solid precursor sources within a tube-furnace architecture. In particular, each precursor should be equipped with its own temperature controller and a dedicated gas delivery line to transport the corresponding vapor species to the substrate, where they can combine at the seed particle to form nanowires. In the case of group IV nanowires, such as Si, growth in a simple tube furnace often results in irregular morphologies with mixed facets and, in some cases, unintended core-shell structures composed of $SiO_x$-Si. By contrast, when more sophisticated vapor delivery systems with separated atomic sources and higher vacuum levels are employed, for example in PLD or MBE, VLS-grown Si nanowires exhibit quality and structural control comparable to those obtained using molecular-precursor-based MOCVD. This comparison suggests that high-quality oxide, carbide, and chalcogenide nanowires may also be achievable through improved synthesis reactor design, even when relying on atomic precursors.

In summary, we identified two principal obstacles in precursor creation and delivery for the non-conventional nanowires: limited availability of suitable molecular precursors and inadequate control over atomic precursor delivery. These constraints matter most through their downstream effect. The accessible precursor chemistries and reactor temperature profiles define which seed compositions can be formed and stabilized during growth, and therefore how much of the seed design space can be explored at all. In this sense, precursor limitations act as an upstream condition on the seed particle behavior discussed in Section 2, rather than as an independent barrier. We anticipate that the identification of appropriate molecular precursors and/or the development of improved synthesis reactors with independent temperature control and separated atomic precursor delivery will widen this design space and enable non-conventional nanowire synthesis to approach the level of control achieved in group IV (Si, Ge) and III-V (GaAs, InAs) systems. We also note that emerging strategies such as salt-assisted VLS synthesis could potentially provide more freedom in creating and delivering the atomic precursors, narrowing the synthesis gap between the conventional and non-conventional nanowires.

## 2. Seed particles

### 2.1 Background

In this section, we discuss the next step in a VLS growth process, namely the phenomena occurring at seed particles and how different seed particles affect nanowire growth kinetics (Fig. 3). We note that the two terminologies, "catalyst particle" and "seed particle," are used interchangeably in the literature[64]. In general, under appropriate conditions, the particle acts as a preferential site for adsorbing precursor species from the vapor phase (Figs. 3a-b) due to its higher sticking probability compared to other available surfaces, such as the substrate or the sidewalls of the nanowire[5]. In this sense, the seed particle serves as a sink that collects material vapors and transports them to the nucleation site, namely the particle-substrate (L-S) interface[46] (Fig. 3a). The chemistry and dynamics of seed particles have shown to control the nanowire growth orientation (Fig. 3c) and crystal phase selection (Figs. 3d-e). During the nanowire growth, the seed particle can exist in either a solid or liquid state, corresponding to the vapor-solid-solid (VSS) or vapor-liquid-solid (VLS) growth mechanisms, respectively. In this review, we focus primarily on the latter, VLS, in which the seed particle forms a eutectic or pseudo-eutectic composition with at least one of the components of the nanowire.

There are two general requirements for an effective seed particle in a VLS growth. First, the seed material should not form strong compounds with any of the nanowire constituents and ideally should not incorporate into the nanowire during growth. Second, the liquid particle (i.e., droplet) must exhibit appropriate wetting behavior and phase stability[13,65,66]. With respect to the former, the seed particle is desired to remain on the top facet of the nanowire throughout the growth process (Fig. 3b). For phase stability, equilibrium phase diagrams are commonly used to identify thermodynamically stable phases under given conditions of temperature, pressure, and composition. However, kinetic factors become increasingly important near the eutectic transition[44], including the stabilization of supercooled liquid droplets at temperatures well below the eutectic temperature (TE)[5], or the coexistence of liquid and solid phases within the same seed particle under certain growth conditions in Au-seeded Si nanowires[67]. In the following sections, we review different classes of seed particles, grouped according to the strength of their interaction with the nanowire materials, with an emphasis on those employed in the growth of non-conventional nanowires.

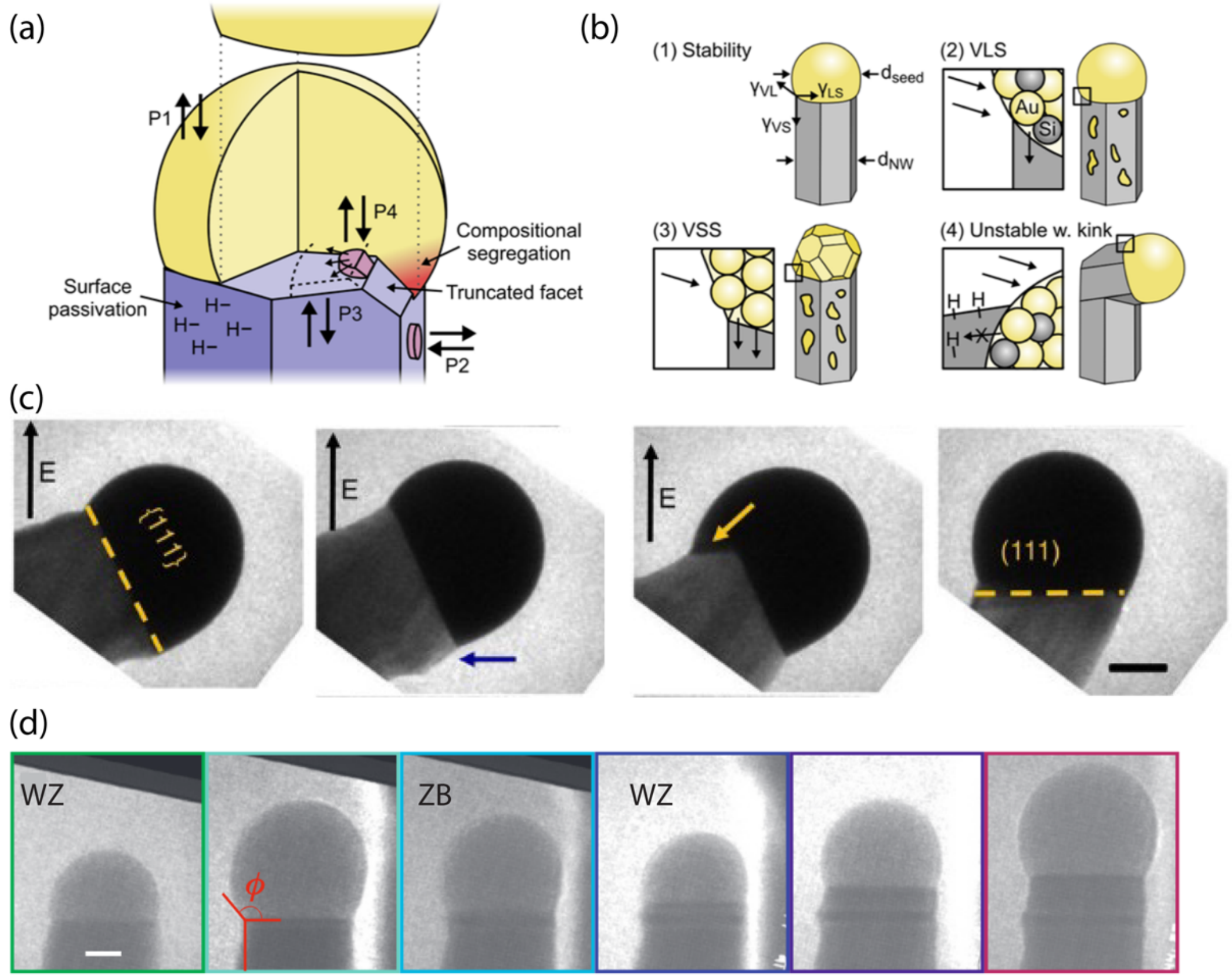

**Figure 3.** Seed particles in VLS synthesis. (a) Precursor fluxes transport to and diffuse within the seed droplet leading to different nucleation sites[46]. (b) Droplet instability, including metal diffusion to the nanowire sidewall, competing VSS nucleation and growth, and droplet roll-off leading to changes in growth directions (i.e., kink nanowires)[5,46]. (c-d) Droplet dynamics and its effects on the nucleation and growth of nanowires. (c) Au-Si's droplet geometry can be tuned by applied electrical bias to *in situ* manipulation of the growth orientation[68]. (d) Au-Ga-As droplet's volume and shape was changed by partial pressure of Ga molecular precursors leading to nucleation of two distinct crystal phases (wurtzite vs. zincblende)[8]. (a-b) Adapted with permission from ref[46]. Copyright 2018 American Chemical Society. (c) Adapted with permission from ref[68]. Available under a Creative Commons CC-BY 4.0 license. Copyright The Authors. (d) Adapted with permission from ref[8]. Copyright 2016 Nature Portfolio.

## 2.2 Weakly interacted seed particles

The first group of seed particles exhibits relatively weak interactions with the nanowire materials. These seeds do not form strong compounds with the nanowire constituents and ideally do not incorporate into the nanowire during growth. Examples include metals, metal alloys, semiconductors, and alkaline-metal compounds (Fig. 4). Table 1 lists the seed materials reported for each class.

Metals are the most commonly used seed materials for synthesis of a wide range of nanowires, spanning the conventional group IV (Si, Ge) and III-V (GaAs, InAs) systems as well as nanowires of oxides, carbides, and chalcogenides (Figs. 4a-b). Among different metals, Au is often regarded as a "universal seed material," enabling the synthesis of a broad variety of nanowires. There are three main reasons underlying this role. First, Au forms eutectic droplets with many materials at relatively low temperatures and with reasonable compositions. For example, the eutectic phase of Au-Si occurs at 363°C and 19.5 at% Si, while that of Au-Sn occurs at 280°C and 29.0 at% Sn. The formation and stability of a liquid droplet are among the key requirements for VLS growth, as discussed earlier. Second, Au is chemically inert with respect to many nanowire materials, such that the droplet primarily acts as a sink to absorb and transport growth species rather than consuming them to form stable intermetallic compounds, which would otherwise suppress growth or introduce unwanted byproducts. Finally, Au can exhibit catalytic activity toward certain molecular precursors, promoting their preferential dissociation on the droplet surface relative to other available surfaces and thereby sustaining VLS growth over competing modes such as non-catalytic vapor-solid growth. In addition to Au, other metals with similar characteristics have been used as VLS seed materials, including Ag, Al, Cu, Pt, and Fe[14,69]. In some cases, alloy seed particles composed of multiple metals are employed when no single metal satisfies all the requirements for stable VLS growth[5].

Despite their effectiveness, metal seed particles also present notable drawbacks. A major concern is the diffusion of metal atoms from the seed particle into the nanowire sidewalls[70] or, in some cases, into the nanowire bulk[71] (Fig. 3b). Such unintended decoration or doping can degrade device performance, as observed in Si-based electronics[71,72] and ZnO photonic devices. To mitigate these issues, alternative weakly interacting, non-metal seed particles have been explored. One example is the use of semiconductor seed particles, most notably Ge[73,74] (Fig. 4c). Ge has been employed to seed the growth of several oxide nanowires, including ZnO[73,74], $GeO_2$[75], mixed $SnO_2$-$In_2O_3$[76], $SiO_2$[77], and $Al_2O_3$[77]. A characteristic feature of Ge-seeded nanowires is their nano-matchstick morphology, in which a relatively large Ge particle resides atop a slender nanowire. This is in contrast to the comparable dimensions between a seed particle's and nanowire's diameter typically observed in metal-seeded growth. However, the applicability of Ge seeds remains limited to a small subset of metal oxides, likely due to the difficulty in forming stable eutectic droplets between Ge and many nanowire materials. Furthermore, recent studies indicate that under certain conditions Ge seeds can diffuse into and react with the nanowire, leading to a competition between growth and etching at the seed-nanowire interface[74].

Another class of weakly interacted, non-metal seed materials comprise alkaline-metal compounds[78], such as Na- and K-based salts (Figs. 4d-e). These materials have been used uniquely for the growth of quasi-one-dimensional structures of transition metal oxides and chalcogenides, including $MoO_3$[79], $WO_3$[80], $WO_3$-$WSe_2$[81], $MoS_2$[82], and $MoSe_2$[83]. Commonly used metal or semiconductor seeds, such as Au, Ag, Ge, are often ineffective for these systems, in part due to the extreme conditions required to form eutectic phases between high-melting-point transition metals (e.g., Mo, W, Ta, Nb) and these seed materials. On the other hand, alkaline metal halides (NaCl, KCl, KI) or hydroxides (NaOH, KOH), when combined with transition metal oxide precursors, can enhance precursor volatility and simultaneously serve as seed particles for nanowire growth[78,82]. At elevated temperatures (600-800°C), reactions between the salt and precursor generate volatile intermediate compounds (e.g., $Na_2$ $MoO_4$), which are transported to the substrate where alkaline-containing droplets are present. Nanowires of transition metal oxides or chalcogenides are then formed via a VLS-like process:

$$NaCl_{(s)} + MoO_{3(s)} \rightarrow Na_2MoO_{4(v)} + Cl_{2(v)} \quad (10)$$
$$Na_2MoO_{4(v)} \rightarrow MoO_{3(nanowire)} \quad (11)$$
$$Na_2MoO_{4(v)} + S_{(g)} \rightarrow MoS_{2(nanowire)} \quad (12)$$

It is worth noting, however, that direct experimental evidence for classical binary-eutectic VLS growth using this class of seed materials remains limited. Moreover, because many of these systems involve layered materials such as $MoO_3$ and $WS_2$, diffusion or intercalation of alkali species into the van der Waals gaps of the nanowires may occur, further complicating interpretation of the growth mechanism.

Beyond enhancing precursor volatility, increasing evidence suggests that alkaline metal halides can actively participate in the catalyst droplet itself, placing these systems within a salt-assisted VLS[58] framework rather than a purely transport-limited or vapor-solid regime. Post-growth compositional analyses have revealed the presence of alkali and halide species within the seed particle together with transition metal and chalcogen components, indicating the formation of multicomponent, alloyed liquid droplets under the growth conditions. In this picture, the salt plays a dual role: it facilitates the generation of volatile metal-containing intermediates and modifies the thermodynamic and kinetic properties of the catalyst droplet, including its effective melting temperature, solubility limits, and interfacial energetics. This coupling between precursor activation and catalyst chemistry distinguishes salt-assisted VLS growth from conventional metal-seeded VLS and from salt-assisted vapor-solid growth.

In summary, weakly interacted seed particles encompass a diverse set of materials, including metals, semiconductors, and alkaline-metal compounds, and enable the growth of a wide range of oxide, carbide, and chalcogenide nanowires, including systems containing high-melting-point metals (see Table 1). Nevertheless, these approaches are not without limitations, with weak but unavoidable interactions between the seed and the nanowire often affecting structural and functional properties.

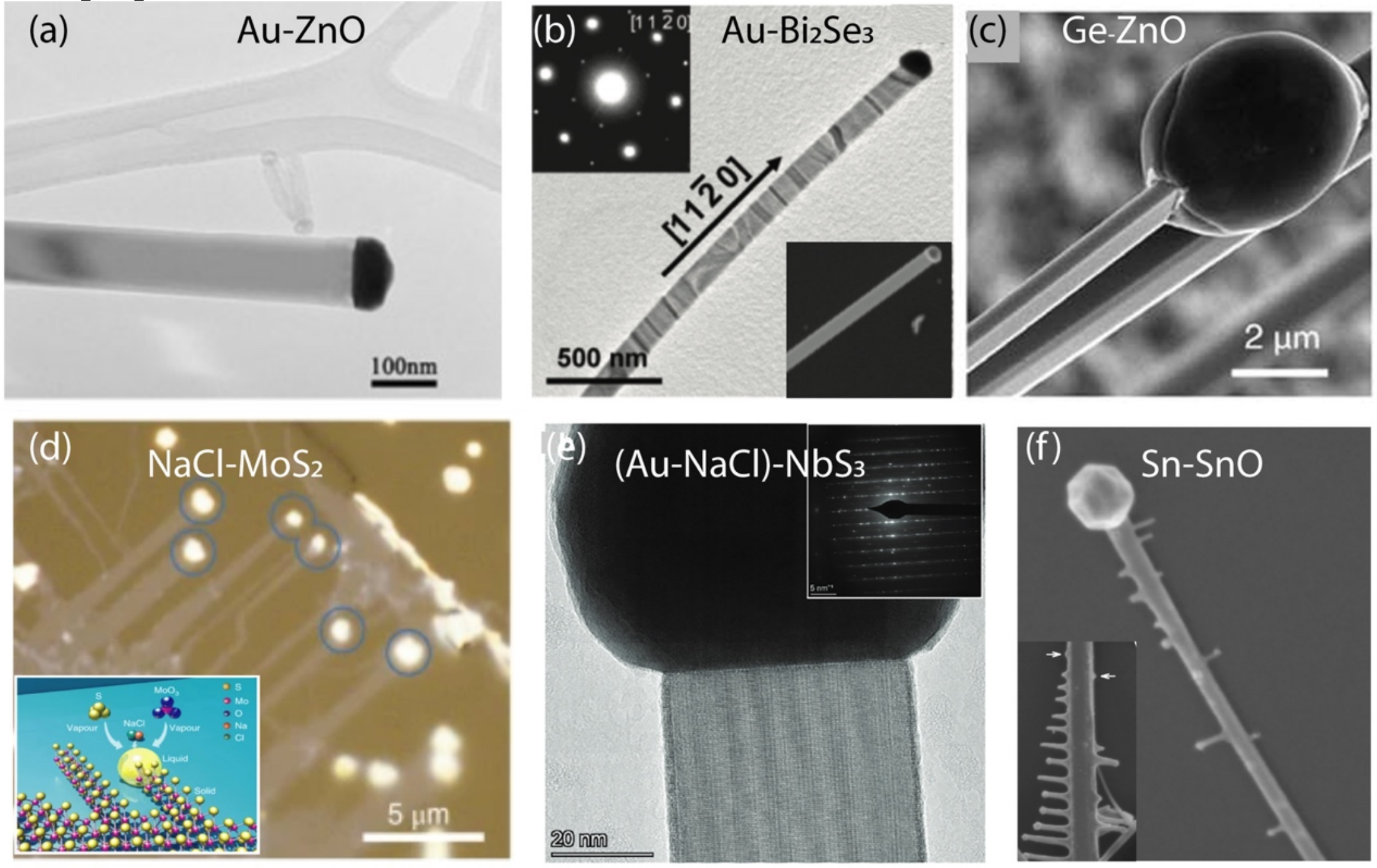

**Figure 4. Chemistry of seed particles for non-conventional nanowire growth.** (a) Au-seeded oxide (ZnO) nanowires[84]. (b) Au-seeded a layered metal chalcogenide nanowire ($Bi_2Se_3$)[30,32]. (Inset) An electron diffraction showing the hexagonal structure of $Bi_2Se_3$ (top) and a SEM image of the nanowire (bottom). (c) A semiconductor (Ge)-seeded a transition metal oxide (ZnO) nanowire[73,74]. Ge particle has an unusually large volume that is 4-6 times bigger than the diameter of the ZnO nanowires. (d) Alkaline metal halide salt (NaCl) seeded $MoS_2$ 1-D nanoribbons. NaCl particles have a higher contrast in the optical image. (Inset) Proposed VLS mechanism for the growth of $MoS_2$ ribbons[82]. (e) Au-NaCl alloy seeded $NbS_3$ nanowires[58]. (f) Self-seeded nanowires (Sn self-seeded SnO nanowires)[85]. (Inset) Arrays of SnO nanowires growing on the sidewall of the main branch due to the seed particle (Sn) diffusion and self-seeded growth. (a) Adapted with permission from ref[84]. Copyright 2002 Wiley-VCH. (b) Adapted with permission from ref[30]. Copyright 2010 American Chemical Society. (c) Adapted with permission from ref[74]. Copyright 2020 IOP Publishing. (d) Adapted with permission from ref[82]. Copyright 2018 Nature Portfolio. (e) Adapted with permission from ref[58]. Available under a CC-BY 4.0 license. Copyright The Authors. (f) Adapted with permission from ref[85]. Copyright 2006 American Chemical Society.

### 2.3 Strongly interacted seed particles

In contrast to weakly interacted seeds, strongly interacted seed particles exhibit strong interactions with at least one component of the nanowire or may even constitute part of the nanowire itself. This situation typically arises when one component of the nanowire is a low-melting-point metal such as Pb, Ga, In, Sn, or Ge (Fig. 4f). In such cases, the metal can simultaneously serve as the seed particle and as a constituent of the nanowire, leading to self-seeded growth. The growth dynamics in these systems are governed by the balance between the inward flux of atoms entering the droplet from the vapor phase and the outward flux of atoms leaving the droplet to incorporate into the growing nanowire[86]. When the inward flux dominates, the droplet volume increases over time, potentially altering droplet geometry, contact angle at the trijunction, and even triggering the nucleation of new crystal phases or droplet instability. Conversely, when the outward flux exceeds the supply rate, the droplet gradually shrinks, often resulting in tapered nanowire morphologies.

Compared to weakly interacted seed particles, strongly interacted or self-seeded growth offers several advantages. Most notably, issues related to seed diffusion and unintentional doping are significantly reduced when the seed material is identical to the nanowire material. Additionally, post-growth removal of the seed particle, which can degrade crystallinity in some group IV and oxide nanowires, is no longer required. However, the applicability of strongly interacted seeds is limited to a relatively small set of materials (Fig. 4f), primarily compounds of low-melting-point metals such as PbS, $Ga_2O_3$, $In_2Se_3$, SnO, and $GeO_2$. Furthermore, this approach restricts the range of accessible heterostructures, even in the conventional nanowire systems, since compositional modulation is limited to elemental substitutions.

### 2.4 Designing *in situ* growth experiments for non-conventional nanowires

Much of the mechanistic understanding summarized in this review for group IV and III-V nanowires was obtained by watching growth happen. *In situ* TEM established the physical state of the droplet, revealed supercooling and two-phase coexistence[44,67], resolved individual nucleation events at the trijunction, and produced the quantitative kinetics that underpin numerical growth models[5,42,46] (Figs. 1c-d). Representative studies are discussed in Section 3.1.2, and dedicated reviews cover this literature in depth[5,46]. On the other hand, limited body of work exists for oxides, carbides, or chalcogenides (Fig. 5). The reasons are specific and technical, and we examine them here because they define what a workable experiment for these materials must overcome.

The first obstacle is gas compatibility. The precursor chemistries described in Section 1 conflict directly with the requirements of an electron microscope column. Metal-organic precursors for transition metals, diethylzinc among them, adsorb strongly on internal surfaces and degrade the vacuum level of the instrument. Chalcogen sources such as $H_2S$ and $H_2Se$ are toxic and corrosive, and their residues are difficult to pump away. Since most high-end instruments operate in shared facilities and national laboratories, the decision to admit such gases is constrained as much by facility safety policy and the risk to a multi-user instrument as by the physics of the experiment. Open-column environmental TEM, the platform that produced the Si and GaAs results[42], is therefore restricted to a narrow set of benign gases that excludes most of the chemistry in Section 1.

Closed gas cells relax the column constraint by confining the reaction volume between electron-transparent membranes, and cells of this type are commercially available. Their track record, however, is built almost entirely on $H_2$, $O_2$, $H_2O$, and light hydrocarbons. How chalcogen vapors, metal-organic species, or halide-containing intermediates interact with SiN membranes, sealing components, and the gas delivery lines of these holders is essentially uncharacterized, and delivering a low-volatility species into a cell at a controlled flux is an unsolved problem in its own right. We regard this less as a barrier than as an open design space. The hardware exists, and what is missing is the compatibility knowledge that the thin-film and catalysis communities long ago accumulated for their own chemistries.

A complementary strategy avoids gas delivery entirely by generating vapor from a solid source inside the cell or on the heater membrane (Fig. 5a). A single-source precursor containing all elements of the target material makes this particularly clean, and the approach has been demonstrated for layered two-dimensional materials, where thermolysis of drop-cast ammonium tetrathiomolybdate or ammonium tethrathiotungstate produced $MoS_2$ or $WS_2$ domains whose nucleation, reorientation, and coalescence were followed *in situ*[87–89] (Fig. 5a). Equivalent single-source chemistries do not yet exist for most nanowire materials, but the demonstration establishes that local solid-source delivery is a viable route. Precursor activation strategies that lower the required temperatures, including the salt-assisted reactions of Section 1.2[58], may widen the set of materials accessible this way.

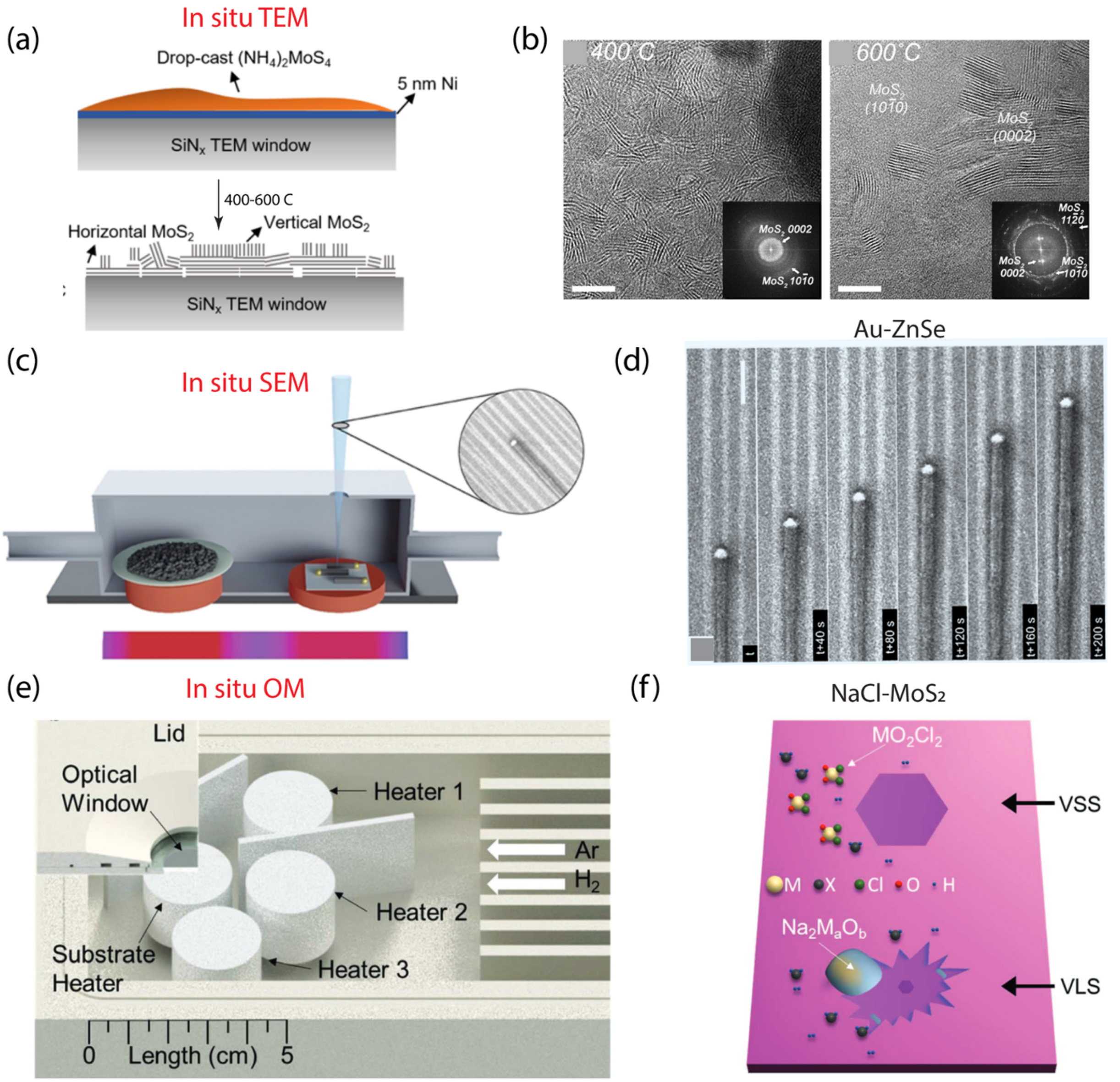

**Figure 5. *In situ* growth experiments**. (a-b) *In situ* TEM: transition metal chalcogenide precursor (ammonium tetrathiomolybdate) is dropped-cast directly on a heating SiN membrane deposited with seeding metal Ni. *In situ* heating experiment shows the formation of vertically grown $MoS_2$ flakes on Ni seed[88]. (c-d) *In situ* SEM showing the VLS nucleation and growth of ZnSe seeded by Au particles[90]. (e-f) *In situ* optical microscopy: home-built furnace compatible with optical microscopy allows real-time observation of salt-assisted $MoS_2$ growth[91]. (a-b) Adapted with permission from ref[89]. Copyright 2019 American Chemical Society. (c-d) Adapted with permission from ref[90]. Copyright 2022 American Chemical Society. (e-f) Adapted with permission from ref[91]. Copyright 2019 Royal Society of Chemistry.

Electron-beam generation of growth species is the workaround used in the few existing *in situ* studies of non-conventional systems. In the sapphire and ZnO experiments discussed in Section 3.1.2, the beam liberated growth species from the substrate itself[92,93], and post-growth heating experiments on Au-ZnO likewise relied on the beam-heated environment rather than delivered precursors[94,95]. These studies show that observation is possible, but the method offers no control over precursor flux, and the beam can remove material from the growing structure at the same time as it supplies it. Growth and etching proceed together, and separating the intrinsic mechanism from beam-driven processes is difficult. Beam-generated species are best viewed as a demonstration of feasibility rather than a foundation for systematic mechanistic study.

Electron microscopy is also not the only path to direct observation. Environmental SEM tolerates higher pressures and cruder vacuum than TEM, and real-time SEM observation of nanowire growth from heated solid sources has been achieved for ZnO[96] and for surface-guided ZnSe nanowires, where the Au droplet was followed as the wire elongated[90]. The costs are spatial resolution, which excludes the interface-level detail that makes *in situ* TEM decisive, and cleanliness, since hydrocarbon deposition under poorer vacuum can alter adatom diffusion and droplet dynamics, the very processes under study. Optical probes with purpose-built reactors extend the same idea to full CVD conditions, and have been used to follow monolayer $MoS_2$ growth in real time[91,97], although the diffraction limit restricts them to structures far larger than a seed droplet. Each platform trades observable detail against chemical realism, and no single one currently offers both.

An experiment capable of resolving the seed particle questions that Section 2.5 sets out therefore requires several elements at once. It must deliver or generate a controlled flux of species that are hostile to conventional instruments, at temperatures matching real growth windows, in an environment clean enough that the observation does not perturb droplet behavior, and with resolution sufficient to see the droplet interior and the liquid-solid interface. None of the current platforms satisfies all four requirements. Progress is likely to come from closed cells with chemistry-specific compatibility engineering, from single-source solid precursors, and from correlating lower-resolution real-time probes with post-growth atomic-scale analysis.

**2.5 Challenges and opportunities**

In this section, we discussed two groups of seed particles, classified by the strength of their interaction with the nanowire material. Weakly interacting seeds, including metals, semiconductors, and alkaline-metal compounds, have produced the widest range of oxide, carbide, and chalcogenide nanowires, at the cost of unintended seed incorporation that can compromise structure and properties. Strongly interacting or self-seeding growth removes the foreign seed but is confined to compounds of low-melting-point metals and restricts compositional modulation. Neither group currently offers what the binary phase diagram provides for conventional systems, namely a predictive basis for selecting a seed and anticipating its behavior.

We regard the seed particle, its material choice, its chemical composition during and after growth, and its dynamics, as the principal bottleneck restricting deterministic VLS synthesis of non-conventional nanowires. The assessment rests on how much remains unknown at this step compared with the others. The physical state of the seed during growth is unresolved in even the most studied systems. For Au-seeded ZnO, growth is assigned to VLS on the grounds that the Au-Zn eutectic lies well below typical growth temperatures, while post-growth heating experiments in which Au particles remained solid up to 1000°C support a VSS assignment instead[94,95]. Whether Au acts catalytically toward precursor dissociation in these systems, as it does for silane, or serves only as a collection site is similarly untested. Droplet compositions are inferred from post-growth analysis rather than measured during growth, an inference that becomes more fragile as droplets become multicomponent. And the factors governing whether a seed remains at the tip, rolls off, or diffuses into the sidewalls, which set tapering, kinking, and unintended doping as described in Sections 2.2 and 3.1.1, are known only empirically. These questions concern the state of the seed rather than the supply of material, and no amount of precursor development answers them.

The gap persists because the experiments that would close it are the ones Section 2.4 describes, and they have not yet been performed for these material classes. This is also why we consider the seed problem an opportunity rather than a dead end. The emergence of multicomponent droplets, in which alkali and halide species alter effective melting temperatures, solubility limits, and interfacial energetics, expands the seed design space beyond binary systems and may itself ease the path to observation by lowering required growth temperatures. The consequences of seed behavior for morphology, defect content, and heterostructure formation are developed in Sections 3 and 4, and we return to the bottleneck identified here in the Summary and Outlook.

## 3. One-dimensional structures and their growth mechanisms

After precursors arrive at the seed droplet, the subsequent steps in the VLS growth process involve crystal nucleation at the droplet-solid (LS) interface and the evolution of the crystal into distinct one-dimensional morphologies. In this section, we begin by discussing the most common one-dimensional morphology, namely the nanowire, and its growth mechanism in both conventional (Si, Ge, GaAs) and non-conventional (ZnO, SnO, $Eu_2O_3$, $Sm_2O_2$, SiC, GaS, GeS) material systems (Fig. 6). Other morphologies, including nanoribbons, twisted nanowires, nanotubes, and heterostructures, can be regarded as derivatives of the nanowire growth process and are discussed subsequently.

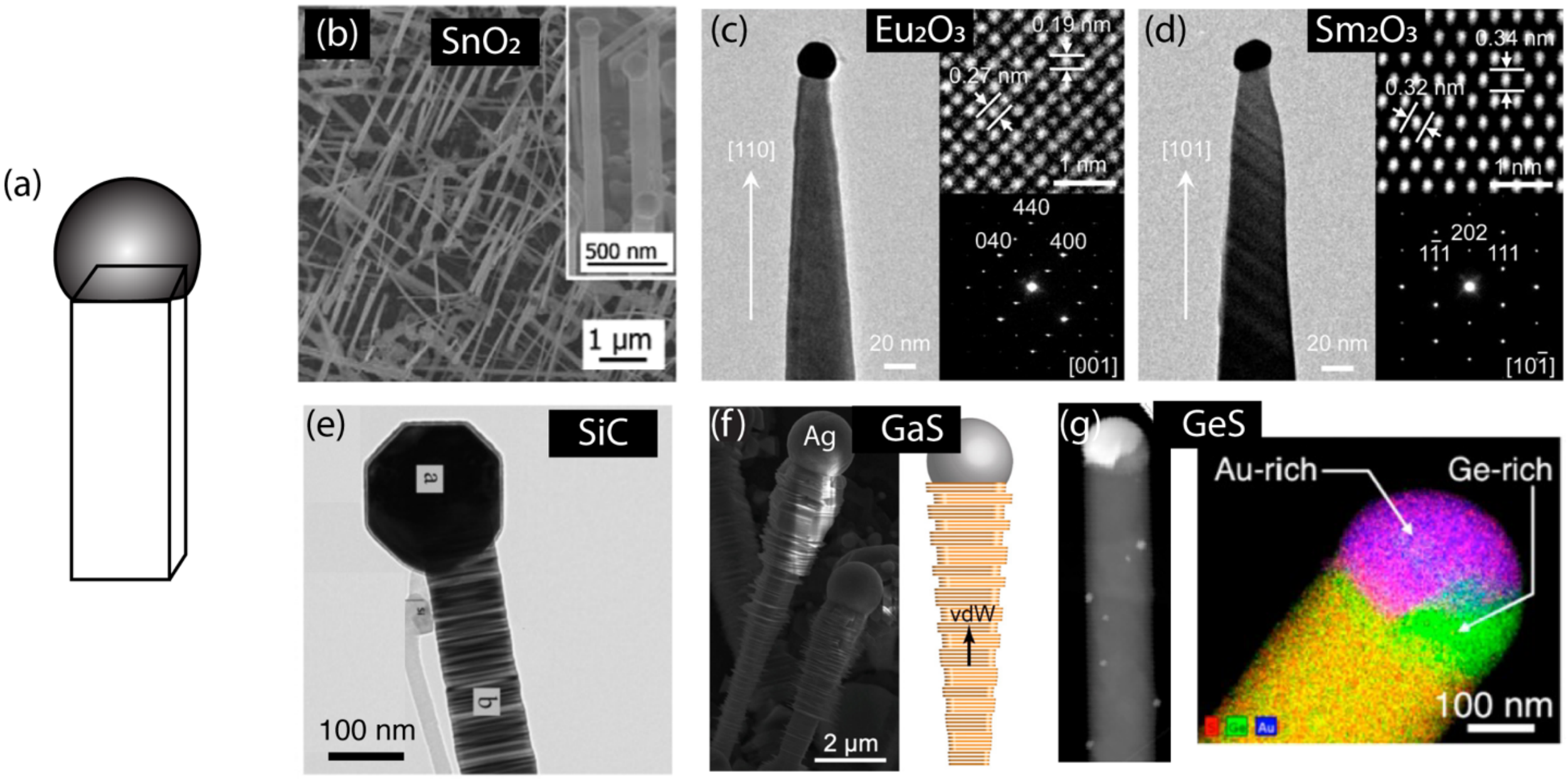

**Figure 6. VLS-grown nanowires.** (a) Schematics showing the telltale VLS-grown nanowire structure: a hemisphere particle sitting atop square-cross-section wire. (b-d) Au-seeded oxide nanowires[15]. (e) Fe-seeded SiC nanowire[23]. (f) Ag-seeded GaS nanowires[53]. (g) VLS growth of layered nanowire GeS[55]. EDS mappings show Ge segregation within the Au-Ge droplet. (b-d) Adapted with permission from ref[15]. Copyright 2015 American Chemical Society. (e) Adapted with permission from ref[23]. Copyright 2007 American Chemical Society. (f) Adapted with permission from ref[53]. Copyright 2020 American Chemical Society. (g) Adapted with permission from ref[55]. Copyright 2020 American Chemical Society.

### 3.1 Nanowires

#### 3.1.1 Structures of nanowires

Nanowires represent the most typical morphology in the family of VLS-grown one-dimensional materials[98], characterized by the presence of a seed particle at the tip and a very high aspect ratio, defined as a ratio of length/diameter or length/width, typically on the order of $10^2$-$10^3$ (Fig. 6a). The nanowire diameter is usually comparable to the size of the seed particle, reflecting the role of the droplet in confining crystallization and directing axial growth. Followed the successful demonstration of VLS-enabled synthesis of conventional nanowires (Si, Ge, GaAs), there are many reports showing the telltale evidence of VLS nanowires in oxides (Figs. 6-d, see also Figs. 4a, c and f), carbides (Fig. 6e) and chalcogenides (Figs. 6f-g, see also Figs. 4b and e).

Two commonly observed derivatives of the ideal nanowire morphology are tapered and kinked nanowires[13,66] (Figs. 6c-d), both of which originate from instabilities of the seed particle droplet. In tapered nanowires, the volume of the seed particle gradually decreases during growth, potentially due to diffusion of seed material onto the nanowire sidewalls (Fig. 6e). Because the droplet size governs the nanowire diameter, a shrinking droplet produces a nanowire with a progressively decreasing diameter toward the tip. In contrast, kinked nanowires arise when the seed droplet migrates away from the center of the nanowire, often toward a corner or facet of the top surface. This displaced droplet can nucleate a new growth segment with a different crystallographic orientation, resulting in a kinked morphology. We note that recent *in situ* spectroscopy studies of Au-seeded Ge and Si nanowires have demonstrated that droplet stability can be actively modulated through surface chemistry, for example by decorating nanowire sidewalls with organic moieties. Such control enables the synthesis of non-tapered nanowires[99], modulation of nanowire diameters[100], and steering of growth directions, thereby allowing systematic control over nanowire cross sections[7]. In the next sections, we will discuss the growth mechanism of nanowire morphology informed by both *in situ* and *ex situ* studies.

#### 3.1.2 Nanowire growth mechanisms revealed by *in situ* studies

Au-seeded Si nanowires represent the first system in which VLS growth was captured quantitatively using *in situ* TEM (Fig. 1c). These observations provided key inputs for numerical models that successfully describe nanowire growth behavior under different temperatures and pressures as discussed in Section 2.4. To illustrate the essential features of VLS growth in a more complex setting, we focus here on Au-seeded GaAs nanowires (Fig. 7a[8]), where crystal phase selection plays a central role. Au-seeded GaAs nanowires can crystallize in either the wurtzite (WZ) or zincblende (ZB) phase depending on growth conditions. The nucleation dynamics and step-flow kinetics of these two phases differ markedly. WZ GaAs typically nucleates under higher V/III (As/Ga) ratios and grows via step flow across the droplet–nanowire interface, with each step corresponding to the addition of a GaAs bilayer and minimal incubation time between successive steps. In contrast, ZB GaAs grows at lower V/III ratios and exhibits oscillatory behavior at the trijunction, manifested by the periodic appearance and disappearance of a truncated top facet. This system highlights the sensitivity of crystal nucleation to droplet geometry, interfacial energetics, and precursor supply, underscoring the value of *in situ* TEM for elucidating growth mechanisms in III-V nanowires. On the other hand, although the VLS mechanism is widely invoked to explain the growth of oxide, carbide, and chalcogenide nanowires, direct experimental evidence remains scarce. Here, we limit discussion to *in situ* TEM studies of two representative systems: self-seeded $Al_2O_3$ nanowires (Fig. 7b) and Sn-seeded ZnO nanowires.

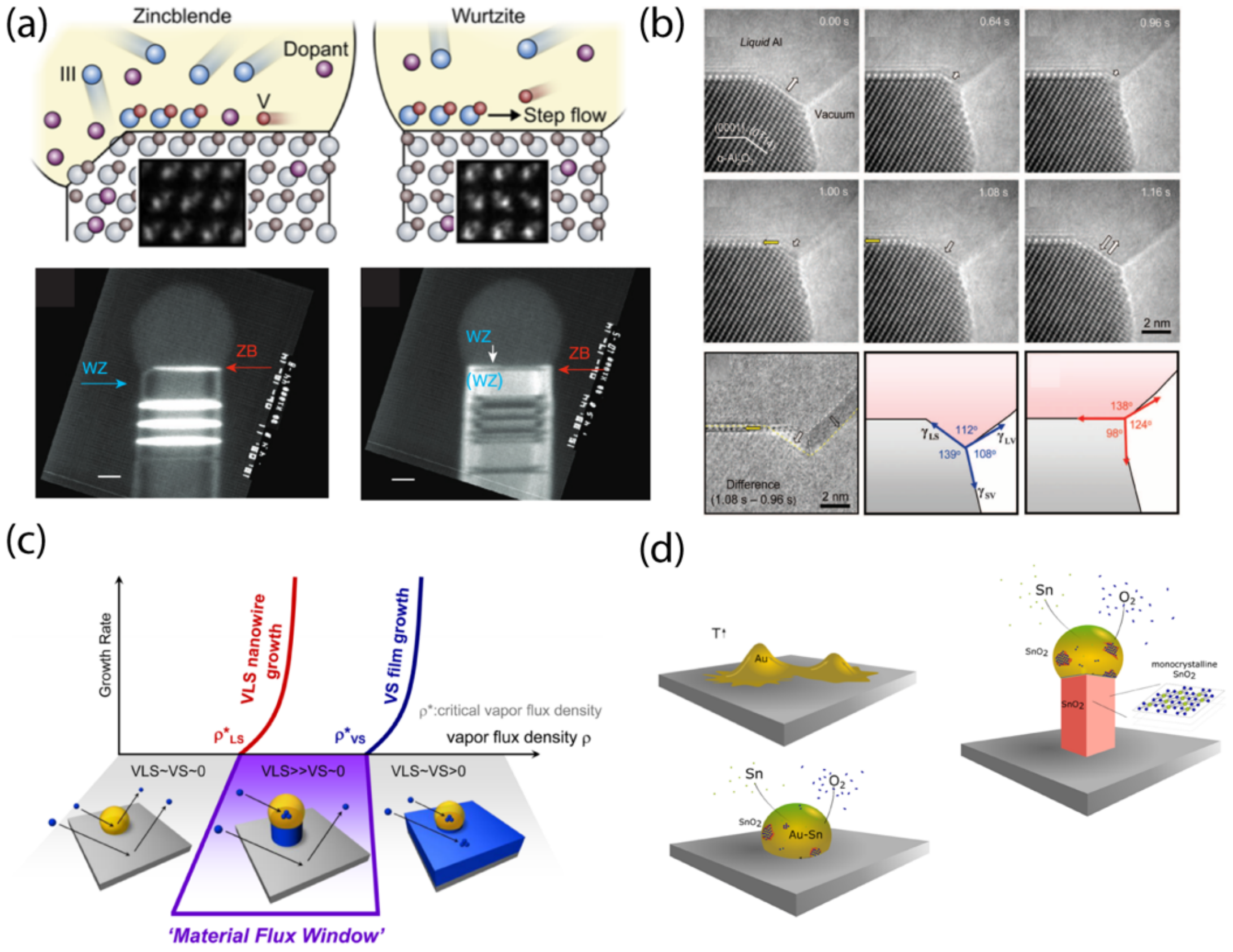

**Figure 7. Growth Mechanism.** (a) VLS growth modes in Au-seeded GaAs nanowire system by *in situ* TEM[8]. The crystal is nucleated at the droplet-nanowire interface following two pathways, depending on the geometry of the seed particle droplet, which result in two distinct crystal phases: zincblende (Left) and wurtzite (Right). (b) Electron irradiation induced growth of $Al_2O_3$ nanowire by self-seed particle[92]. The formation of a new layer of $Al_2O_3$ at the droplet-nanowire interface is accompanied by the mass oscillation of a number of facets at the trijunction. (c) Flux window growth[15,101] in which there exists a range of precursor flux that exclusively favors the nucleation at the liquid-solid (LS) interface, resulting in the growth of a nanowire instead of a thin film. (d) VLS Flake model[47]. The nanowire forming reaction happens on the surface of the seed droplet instead of at the droplet-nanowire interface (conventional VLS). The material flakes float on the surface reaching droplet-nanowire interface and re-arrange into nanowire crystal. (a) Adapted with permission from ref[8]. Copyright 2016 Nature Portfolio. (b) Adapted with permission from ref[92]. Copyright 2010 American Association for the Advancement of Science. (c) Adapted with permission from ref[15]. Copyright 2015 American Chemical Society. (d) Adapted with permission from ref[47]. Copyright 2018 American Chemical Society.

In the first case (Fig. 7b), the growth of self-seeded sapphire (α-$Al_2O_3$) nanowires was examined using an *in situ* TEM setup similar to that employed for Si nanowires[92]. In this experiment, Al or $Al_2O$ vapor was generated by heating the $Al_2O_3$ substrate above 660°C while simultaneously irradiating the sample with a 1250 keV electron beam. Oxygen was likely supplied from residual gases within the TEM column. The resulting $Al_2O_3$ nanowires grew along the [0001] direction and were bounded by (00-12) and (0-114) facets. These facets exhibited periodic oscillations, which facilitated transport of oxygen species from the vapor-liquid (VL) interface to the liquid-solid (LS) interface during nucleation of new α- $Al_2O_3$ (0006) layers. Growth proceeded in a layer-by-layer manner, analogous to observations in Au-seeded GaAs nanowires (Fig. 7a), including the appearance of truncated facets and finite incubation times between nucleation events.

In a second study, the growth of Sn-seeded ZnO nanowires was investigated to understand conditions that stabilize the metastable ZB phase of ZnO[93], in addition to the more commonly observed WZ phase. In this system, growth species were generated *in situ* by 200 keV electron-beam irradiation of a $Zn_2SnO_4$ substrate. Under these conditions, Sn-seeded ZnO nanowires exhibited both WZ and ZB phases, sometimes on different sidewalls of the same nanowire. WZ ZnO consistently nucleated away from the trijunction, forming an edge facet within the liquid droplet, whereas ZB ZnO nucleated directly at the trijunction. The correlation between droplet geometry, nucleation site, and phase selection closely mirrors observations in Au-GaAs systems, reinforcing the notion that key aspects of VLS growth are universal across disparate material systems.

It is important to note that the *in situ* experimental configurations used to study Sn-ZnO and Al-$Al_2O_3$ systems differ substantially from those employed for group IV and III-V nanowires[5], particularly in terms of precursor generation and delivery (see section 2.4 and Fig. 5). In oxide systems, growth species are often generated *in situ* via electron-beam irradiation rather than supplied externally as controlled vapor-phase precursors. Despite these differences, the fundamental growth features observed in both conventional (Si, GaAs) and non-conventional (ZnO, $Al_2O_3$) nanowires are remarkably similar, underscoring the broad applicability of the VLS framework.

### 3.1.3 Nanowire growth mechanisms inferred from *ex situ* studies

In addition to the limited number of *in situ* investigations, several growth mechanisms have been proposed based on *ex situ* post-growth characterization. Here, we discuss two representative mechanisms derived from commonly used synthesis approaches: pulsed laser deposition (PLD) and chemical vapor deposition (CVD) using a horizontal tube furnace setup.

The first mechanism, termed "flux window growth" (Fig. 7c), was proposed by Yanagida *et al.*[15,101]. Using PLD, the authors synthesized a variety of single-crystalline oxide nanowires, including ZnO, $In_2O_3$, SnO, and MgO. Conventionally, precursor flux is assumed to primarily influence growth rate. However, these studies revealed the existence of a specific flux window within which nucleation at the liquid-solid interface is favored over vapor-solid deposition. When growth parameters are tuned to remain within this window, material incorporation proceeds predominantly along the axial direction, suppressing lateral growth on nanowire sidewalls. The width of this flux window was shown to depend on both material chemistry, with wider windows for oxides exhibiting stronger metal-oxygen bonding (Fig. 7c), and temperature. This approach enabled the synthesis of previously inaccessible oxides, such as MnO, CaO, $Sm_2O_3$, NiO, and $Eu_2O_3$, as well as some common oxides (ZnO, $SnO_2$) at substantially reduced temperatures compared to typical physical vaporization methods discussed in Section 1.

A second *ex situ* mechanism, known as the VLS flake (VLSF) model (Fig. 7d), was proposed by Zacharias *et al.*[47] based on studies of Au-$SnO_2$ nanowire growth. In contrast to the conventional VLS picture, where the nanowire-forming reaction occurs exclusively at the droplet-nanowire interface, the VLSF model proposes that reactions occur on the surface of the seed droplet, producing small, solid $SnO_2$ flakes that float on the droplet surface. These flakes migrate along the droplet curvature and are ultimately incorporated at the droplet-nanowire interface, where they reorganize into the crystalline nanowire. Several driving forces have been proposed to facilitate this process, including viscosity gradients arising from compositional inhomogeneity and energetically favorable attachment at the solid interface. This model is supported by observations of flake-like structures coating seed particles in ITO nanowire growth[61].

### 3.2 Other one-dimensional structures and their growth models

#### 3.2.1 Nanoribbons

Nanoribbons, also referred to as nanobelts (Figs. 8a-c), represent a special class of one-dimensional structures characterized by substantially larger widths and typically rectangular cross sections[13,30,102]. As a result, they are often described as quasi-one-dimensional materials. Nanoribbons generally form under two scenarios. First, in materials where the difference in nucleation energy barriers between two crystallographic planes is small, growth can proceed rapidly along one direction while maintaining a non-negligible growth rate along a second direction, producing ribbon-like morphologies. This behavior has been observed in several oxide systems, including ZnO[66], $Ga_2O_3$[51], and SnO[102]. Second, in layered materials that naturally favor plate-like growth, competition between wire-like and plate-like configurations can yield ribbon morphologies, as seen in $Bi_2Se_3$[28,32] and $In_2Se_3$[57] (Figs. 8b-c).

The role of VLS growth in nanoribbon formation remains inconclusive. Post-growth characterization often reveals nanoribbons lacking a seed particle or possessing a seed that is much smaller than the ribbon width. Moreover, seed particles are frequently found at off-center or random locations rather than at the ribbon midpoint. Consequently, two growth mechanisms have been proposed[102]. In the first, nanoribbons initially grow via a conventional VLS process, after which the seed particle rapidly diffuses away from the top surface and diminishes in size. In the second, vapor-solid growth contributes significantly through sidewall incorporation, leading to preferential lateral expansion relative to thickness. This latter mechanism is consistent with the observation that nanoribbons are more prevalent at higher growth temperatures, where non-catalytic vapor-solid deposition is enhanced. Recent reports using salt-assisted VLS growth[58] show clear evidence of $NbS_3$ ribbons with their widths comparable to the diameter of the seeded Au particles.

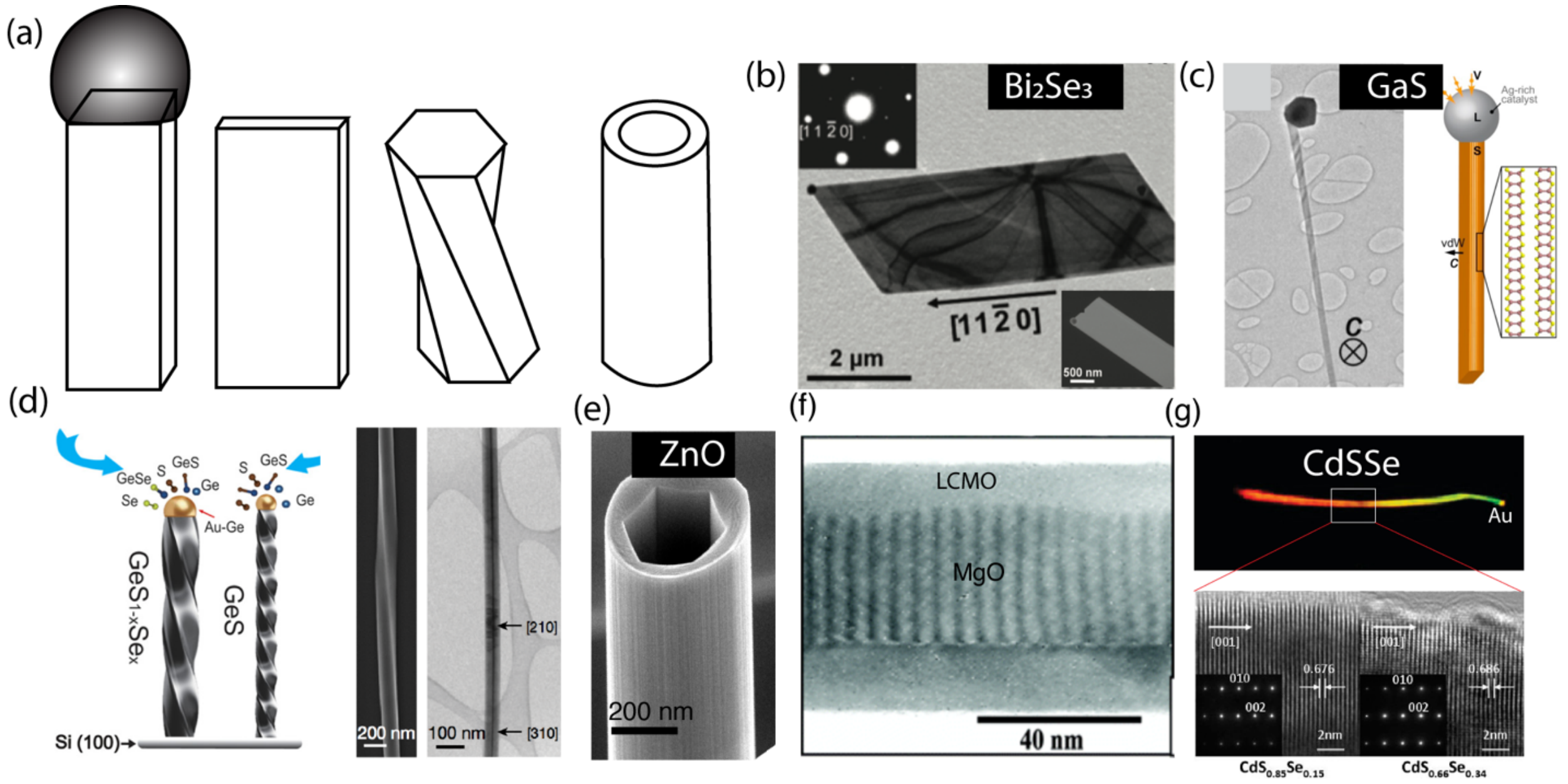

**Figure 8. Other one-dimensional structures.** (a) Schematics showing the most common other one-dimensional nanostructures: nanoribbons, twisted nanowires and nanotubes. The seed particle is illustrated only in the first schematic for simplicity. (b) Au-seeded $Bi_2Se_3$ nanoribbons[30]. (Inset) An electron diffraction showing the hexagonal structure of $Bi_2Te_3$ (top) and a SEM image of the nanoribbon (bottom). (c) Ag-seeded GaS nanoribbons[53]. The seed particle's diameter is much bigger than the thickness of the ribbon. (d) Twisted nanowire of GeS and GeSSe[56]. The twisting motif is clearly displayed in the SEM image. TEM image showing the presence an axial screw dislocation which induces the growth of the twisted structure. (e) Ge-seeded ZnO nanotube[74]. Ge seed particle not shown to display the twisting structure of the nanotube. (f) A core-shell nanowires of MgO and $La_{0.67}Ca_{0.33}MnO$ (LCMO)[15,45]. (g) Alloyed CdSSe by Au seeding[103]. (b) Adapted with permission from ref[30]. Copyright 2010 American Chemical Society. (c) Adapted with permission from ref[53]. Copyright 2020 American Chemical Society. (d) Adapted with permission from ref[56]. Copyright 2019 Nature Portfolio. (e) Adapted with permission from ref[74]. Copyright 2020 IOP Publishing. (f) Adapted with permission from ref[45]. Copyright 2004 American Chemical Society. (g) Adapted with permission from ref[103]. Copyright 2018 Royal Society of Chemistry.

### 3.2.2 Twisted nanowires and nanotubes

Beyond straight nanowires and nanoribbons, VLS-grown one-dimensional materials can adopt twisted morphologies[104,105] (Figs. 8d-e), typically associated with the presence of an axial screw dislocation that induces an Eshelby twist[106,107]. Twisted nanowires often exhibit alternating crystallographic orientations along their length, resulting in a helical structure. The Eshelby twist has been reported in a variety of materials, including PbSe[108], PbS[109], SiC[21,22], ZnO[74,110], and more recently in layered GeS[56,111] and GeSSe[112] nanowires. These twisted nanowires can also serve as templates for higher-order hierarchical structures, as demonstrated in PbS[109], PbSe[108], and GeS[56]. Interest in twisted nanowires has grown considerably since these initial reports, driven by the recognition that an axial screw dislocation in a layered crystal produces a continuous interlayer twist, a synthesis-based analogue of the twisted interfaces otherwise assembled from exfoliated flakes. Control over dislocation character and placement[113], twist-rate modulation through growth chemistry[112], and direct measurements of dislocation-modified transport[114] have followed. We discuss these developments in the context of van der Waals nanowires in Section 4.4.

Substantial evidence supports the coexistence of catalytic VLS growth and screw-dislocation-driven twisting in these systems. However, the detailed crystallization process at the seed particle-nanowire interface in the presence of a screw dislocation remains poorly understood. The origin of the screw dislocation itself is also unresolved, with competing hypotheses proposing nucleation at either the seed particle or the substrate. Notably, some nanoribbons also exhibit twisted morphologies, particularly in ZnO and $SnO_2$ systems. In these cases, twisting arises from electrostatic effects associated with oppositely polarized sidewalls. For example, ZnO nanoribbons possess (0001) and (000-1) surfaces with opposite polarity, and curling of the ribbon minimizes electrostatic energy, producing rings[115] and springs[116]. However, these structures are generally attributed to vapor-solid growth mechanisms.

Another intriguing one-dimensional morphology is the nanotube[117] (Fig. 8e), which has been observed in VLS-grown oxide ($In_2O_3$, ZnO) and carbide (SiC) systems. Nanotube formation is often attributed to the presence of an axial screw dislocation and associated Frank micropipe mechanisms. In some cases, VLS signatures, such as a seed particle at the tip, have been observed in nanotube structures[74]. Nevertheless, the detailed nucleation and growth mechanisms of VLS nanotubes remain largely unexplored. Recently, we proposed an alternative mechanism to explain highly twisted ZnO nanotubes seeded by Ge[74], involving a coexistence of growth and etching at the seed-nanowire interface that ultimately produces a large hollow core.

### 3.2.3 Heterostructures

One of the most compelling features of VLS growth is the ability to incorporate multiple materials or crystallographic phases into a single nanowire to form heterostructures with enhanced or emergent functionalities[118]. This capability has been extensively exploited in group IV (Si, Ge)[10,11] and III-V (GaAs, InAs, GaN, InP)[6] nanowires. In contrast, only a limited number of lateral heterostructures and core-shell nanowires have been reported in oxide systems (Fig. 8f) and chalcogenides (Figs. 8d and g), and there is currently little unambiguous evidence for VLS-based axial heterostructures in oxides, carbides, or chalcogenides.

This disparity likely arises from challenges discussed in earlier sections, particularly limitations in precursor creation and delivery and the poorly understood behavior of seed particles in the non-conventional systems. In particular, the lack of synthesis platforms capable of independently delivering atomic precursors to the growth substrate hinders the ability to switch precursor sources dynamically, which is essential for axial heterostructure formation. These constraints continue to limit the extension of VLS heterostructure concepts from the conventional semiconductor nanowires to more complex material classes.

## 3.3 Challenges and opportunities

In summary, this section has discussed a range of one-dimensional structures realized in the non-conventional materials and their growth mechanisms based on both *in situ* and *ex situ* studies. Despite significant progress in elucidating fundamental growth processes and expanding the accessible morphology space, several key challenges remain. At the same time, these limitations define clear opportunities for advancing the synthesis and integration of one-dimensional nanomaterials. Below, we highlight two major opportunity areas together with their associated challenges and defer opportunities in extending VLS growth to new materials to Section 4.5, where they are discussed together with the emerging one-dimensional systems introduced there.

***Opportunities in growing new structures***

To the best of our knowledge, there is currently no general synthesis protocol capable of producing axial heterostructured nanowires composed of different materials, distinct crystal phases, or dissimilar morphologies in the non-conventional material systems, beyond what has been demonstrated in group IV and III-V nanowires[118]. As discussed throughout this review, this gap arises from multiple factors, most notably the limited understanding of seed particle behavior during nucleation and growth in oxides, carbides, and chalcogenides. Informed by seminal studies on the role of surface chemistry in Si and Ge nanowire growth[7,99,100,119], we hypothesize that improper or uncontrolled sidewall passivation during growth promotes non-catalytic vapor-solid incorporation along nanowire sidewalls rather than exclusive nucleation at the seed particle-nanowire interface. Framed in the terms of Section 2.5, the failure is one of seed exclusivity: the seed particle loses its role as the sole nucleation site, and growth is no longer channeled through the droplet. As a result, many attempts to synthesize heterostructured nanowires in the non-conventional materials yield core-shell architectures rather than true axial heterojunctions[45,62].

At the same time, this challenge presents a substantial opportunity to expand the library of accessible one-dimensional structures. Beyond axial heterostructures composed of different materials or crystal phases, the integration of distinct morphologies within a single one-dimensional object, for example, nanowire-nanoribbon, straight nanowire-twisted nanowire, or twisted nanowire-nanotube heterostructures, may enable new physical phenomena. In practice, materials synthesized under nominally identical conditions often exhibit multiple morphologies within the same growth batch, suggesting that relatively small variations in growth parameters may be sufficient to toggle between different structural outcomes. Controlled access to such morphology-modulated heterostructures could enable emergent functionalities, including modulated luminescence in straight-twisted nanowire junctions[56,111,120] and enhanced thermoelectric performance in twisted nanowire-nanotube heterostructures[114,121,122].

Additionally, nanowire architectures also offer a unique platform for integrating highly lattice-mismatched materials that are difficult or impossible to combine in planar thin-film geometries. Planar heterostructures have already demonstrated striking examples of emergent functionality, such as room-temperature multiferroicity in complex oxides[123,124] and enhanced superconducting transition temperatures at oxide-superconductor interfaces[125]. One-dimensional heterostructures further expand this design space by leveraging their small lateral dimensions and high aspect ratios to more effectively relax interfacial strain. This capability may enable the realization of heterostructures composed of highly dissimilar materials with minimal defect formation[126], opening pathways toward applications in spintronics[127], neuromorphic computing and sensing[128].

***Opportunities in exploring unusual growth modes***

As highlighted in this review, the non-conventional materials exhibit a wide range of one-dimensional morphologies that are rare or absent in group IV and III-V systems, including nanoribbons and twisted nanowires. These observations point to growth modes that are unique to specific material chemistries and crystal structures, in addition to the broadly shared features of the VLS mechanism. Fundamental questions remain open, such as which parameters govern morphological transitions (e.g., nanowire-to-nanoribbon) and how screw dislocations can be introduced or eliminated to switch between straight and twisted growth modes.

Addressing these questions will require advances in both synthesis platforms and characterization strategies. In particular, growth systems capable of independent and rapid control over multiple precursor sources are essential for systematically probing growth kinetics and interfacial processes. Semi-automated, hybrid chemical vapor deposition reactors that integrate inorganic and metal-organic precursor delivery, independent temperature control, and *in situ* precursor switching provide a promising pathway toward this goal. Such platforms enable precise modulation of metal and chalcogen fluxes, total pressure, and growth timing, thereby allowing controlled perturbation of growth conditions during a single experiment. Equally important is the ability to couple advanced synthesis with comprehensive characterization and data-driven analysis. The creation of searchable materials databases that integrate synthesis parameters, structural and chemical characterization, and measured properties can facilitate quantitative identification of the key parameters governing growth outcomes. When combined with semi-automated synthesis, these databases provide the large, high-quality labeled datasets necessary for applying machine learning approaches to materials growth. Data-driven models can be used to rank the relative importance of synthesis parameters, improve reproducibility, and ultimately guide optimization toward targeted structures, defect configurations, and functional properties.

Together, these developments suggest a path toward a more predictive and programmable framework for one-dimensional materials synthesis of unconventional materials. Table 2 summarizes the gaps in synthesis of conventional vs. non-conventional nanowires, grouping by the VLS conditions as discussed in our review (sections 1-3) and highlighting the challenges in controlled synthesis of the non-conventional group. Next, we will apply these delineated rules for a specific group of non-conventional nanowires: emerging van der Waals nanowires and other exotic materials (section 4).

**Table 2.** Comparison of VLS growth between conventional (group IV and III-V) and non-conventional nanowires at each stage of the growth process, summarizing Sections 1 to 3. The seed selection rule, fourth row, is where the two groups differ most fundamentally and is the bottleneck identified in Section 2.5. Section 4 examines how these contrasts apply to van der Waals nanowires, where several of them appear together in a single material family.

| | **Conventional nanowires** Group IV (Si, Ge) and III-V (GaAs, InAs, GaN) | **Non-conventional nanowires** Oxides, carbides and chalcogenides |
|---|---|---|
| **Molecular precursors** (Section 1.1) | Mature library inherited from thin-film deposition: silane, disilane, germane, trimethylgallium, arsine. Broad dissociation windows, high-vacuum compatible | Few suitable candidates. ALD-type precursors behave differently at VLS temperatures; $H_2S$ and $H_2Se$ degrade vacuum systems and cross-contaminate reactors |
| **Atomic precursors** (Section 1.2) | Available (PLD, MBE) with independent, shuttered sources; used when molecular routes are impractical | Dominant route by necessity. Requires activation: carbothermic reduction, hydrogen assistance, or salt-generated volatile intermediates for high-melting metals. |
| **Synthesis reactor** (Section 1.3) | MOCVD, CVD, MBE with independent temperature zones, mass-flow control, fast gas switching | Horizontal tube furnace. Precursor and substrate temperatures coupled through a single gradient; no dynamic flux modulation |
| **Seed selection rule** (Sections 2.1, 2.5) | Binary/ternary phase diagram predicts droplet formation, composition, and stability; seed choice is a design step | No predictive rule beyond binary eutectics. Au used empirically; multicomponent salt-alloyed droplets lack measured phase diagrams |
| **Typical seed materials** (Sections 2.2, 2.3) | Au as near-universal seed; Ag-Au and other alloys for tuning; self-seeding by group III metals | Au where it happens to work; Ge for select oxides; alkali salts for Mo and W compounds; Au-alkali halide alloys for trichalcogenides; self-seeding limited to low-melting-metal compounds |
| **Eutectic accessibility** (Section 2.1) | Low-temperature eutectics at practical compositions (Au-Si: 363°C, 19.5 at% Si) | Frequently inaccessible. Constituent high-melting-temperature metals in trichalcogenides; droplet state |
| | | during oxide growth contested, with Au-ZnO assigned both VLS and VSS [94,95] |
| **Seed particle state during growth** (Sections 2.4, 2.5) | Resolved by *in situ* TEM: liquid versus solid, supercooling, two-phase coexistence directly observed [42,67] | Largely inferred from post-growth analysis and modified binary phase diagrams; direct observation lacking for most systems |
| **Structural control achieved** (Section 3.1) | Deterministic: diameter, position, growth direction, crystal phase (WZ versus ZB), kinking, cross section | Empirical: broad diameter and morphology distributions within single growth runs; phase and orientation rarely selectable, although droplet-size-selected morphologies have been demonstrated in trichalcogenides [58] |
| **Heterostructures** (Section 3.2.3) | Axial junctions with abrupt interfaces (Si-Ge, GaAs-InAs) [10,11]; superlattices; radial shells by design | Radial core-shell structures; no unambiguous VLS axial heterojunctions; single-step mixed-dimensional 1D-2D structures in monochalcogenides [133] |
| **Primary growth instabilities** (Sections 3.1, 3.2) | Droplet-driven tapering and kinking, understood and suppressed through surface chemistry and flux control [7,99,100] | Competing non-catalytic VS deposition on sidewalls; seed diffusion and incorporation; ribbon formation by lateral VS growth; growth-etching competition at the seed interface |
| ***In situ* growth observation** (Section 2.4) | Extensive ETEM and UHV-TEM record for Si, Ge, GaAs [5,43,46]; quantitative kinetics feed numerical growth models | Very few studies; growth species often generated by electron-beam irradiation rather than delivered as controlled vapor [92,93] |

## 4. Van der Waals and other emerging nanowires

### 4.1 Classification of one-dimensional van der Waals materials

Van der Waals materials have appeared at several points in this review, from the layered chalcogenides seeded by metals and salts in Section 2.2 to the nanoribbons and twisted structures of Section 3.2. Here we consider them as a single group and ask how the framework developed in the previous sections, namely precursor creation and delivery, seed particle selection, and the resulting one-dimensional structures, applies to them. All of these materials contain van der Waals gaps, but the structural unit bounded by the gaps is not the same. This difference sets the demands placed on the seed particle and the morphologies that growth can produce, and it therefore organizes the discussion that follows. A dedicated review of one-dimensional van der Waals nanostructures grown by VLS is available[129], and we focus here on what the comparison with the preceding sections reveals rather than on a comprehensive survey of this material family.

We group the materials by composition, consistent with the rest of the review. Group III and group IV monochalcogenides[129], such as GaS, GaSe, GeS, and GeSe (Figs. 9a-c), are built from two-dimensional layers analogous to those of transition metal dichalcogenides. Pnictogen and post-transition metal chalcogenides[30], including $Bi_2Se_3$, $Bi_2Te_3$, and $In_2Se_3$, are built from thicker multi-atom slabs, for example the quintuple layer of $Bi_2Se_3$ (Figs. 9d-e). Transition metal trichalcogenides, $MX_3$ with M = Ti, Zr, Hf, Nb, or Ta and X = S, Se, or Te, consist instead of one-dimensional chains coupled weakly in both transverse directions[125] (Figs. 9f-h). The compositional grouping also tracks the constraints established in Sections 1 and 2. In the first two families, the constituents are volatile at moderate temperatures and the metals form liquid droplets with common seeds, so precursor delivery and seed selection follow conventional routes. In the trichalcogenides, the transition metals melt above 2000°C, supply negligible vapor pressure under typical CVD conditions, and form no accessible binary eutectic with common seed metals. For this family, every classical requirement fails simultaneously. Sections 4.2 to 4.4 discuss each family in turn.

A second distinction concerns the origin of the one-dimensional morphology. Intrinsic structural anisotropy is not a requirement for nanowire growth. Si and Ge are bonded isotropically in three dimensions, yet they form the archetypal VLS nanowires because the droplet confines nucleation to the liquid-solid interface and suppresses incorporation elsewhere, as discussed in Sections 2 and 3. At the opposite limit, the one-dimensionality of a trichalcogenide is encoded in its bonding before any seed particle is introduced. The layered families lie between these limits. Their crystals prefer two-dimensional extension, and a seed particle must work against this preference to produce a wire rather than a plate. The coexistence of wires, ribbons, and plates in $Bi_2Se_3$ and $In_2Se_3$ under similar growth conditions[30], noted in Section 3.2.1, reflects this competition (Figs. 9d-e).

These two axes together define what the seed particle must accomplish in each family, and they reveal a trend that runs opposite to the difficulty of seeding itself. As intrinsic anisotropy increases from the isotropic semiconductors to the chain-like trichalcogenides, the morphological burden on the droplet decreases, while the difficulty of forming a stable droplet increases. The monochalcogenides are simple to seed but require the droplet to enforce one-dimensionality. The trichalcogenides supply their own anisotropy but resisted seeding entirely until multicomponent droplets became available[58]. This inversion explains why the two families followed such different historical trajectories despite their shared van der Waals character, and it frames the discussion of each family in the following subsections.

One further structural variable has no counterpart in group IV and III-V nanowires. In the layered families, the van der Waals stacking direction can lie either along or perpendicular to the growth axis, and the two orientations correspond to different growth modes rather than to a minor morphological detail. We return to this distinction in Section 4.2.

### 4.2 Layered van der Waals materials: monochalcogenides and layered chalcogenides

The monochalcogenides place the fewest demands on precursor delivery of any material class in this review. GaS, GaSe, GeS, and GeSe evaporate congruently or near-congruently at temperatures accessible in a standard tube furnace, and growth from the resulting vapor requires none of the carbon-, hydrogen-, or salt-assisted activation schemes of Section 1.2[53–55]. Seed selection is likewise classical. Au, Ag, and Bi all form liquid droplets with the group III and group

IV metals at moderate temperatures, and growth proceeds well below 500°C. By the criteria of Sections 1 and 2, these are among the best-behaved VLS systems outside the conventional semiconductors[129].

The characteristic difficulty appears at the third step, the evolution of the one-dimensional structure. A layered crystal prefers to extend laterally, and the droplet must confine growth to one dimension against that preference. The outcome of this competition is registered in the orientation of the van der Waals stacking relative to the growth axis. When the layers lie perpendicular to the growth axis, the wire elongates by repeated nucleation of new layers at the liquid-solid interface, a mode directly analogous to the layer-by-layer growth of GaAs described in Section 3.1.2. When the layers lie parallel to the growth axis, the full stack is established at the start of growth and the structure extends by incorporation at the layer edges. Edge-fed growth yields ribbons of uniform thickness, and because no new layer nucleation occurs during elongation, the stacking registry set at nucleation persists along the entire ribbon[131]. Ag-seeded GaS nanowires exhibiting both layer orientations have been reported from the same growth chemistry[53] (see Figs. 6f and 8c), indicating that orientation selection occurs at the nucleation stage and is sensitive to the seed and its interface rather than to the precursor supply.

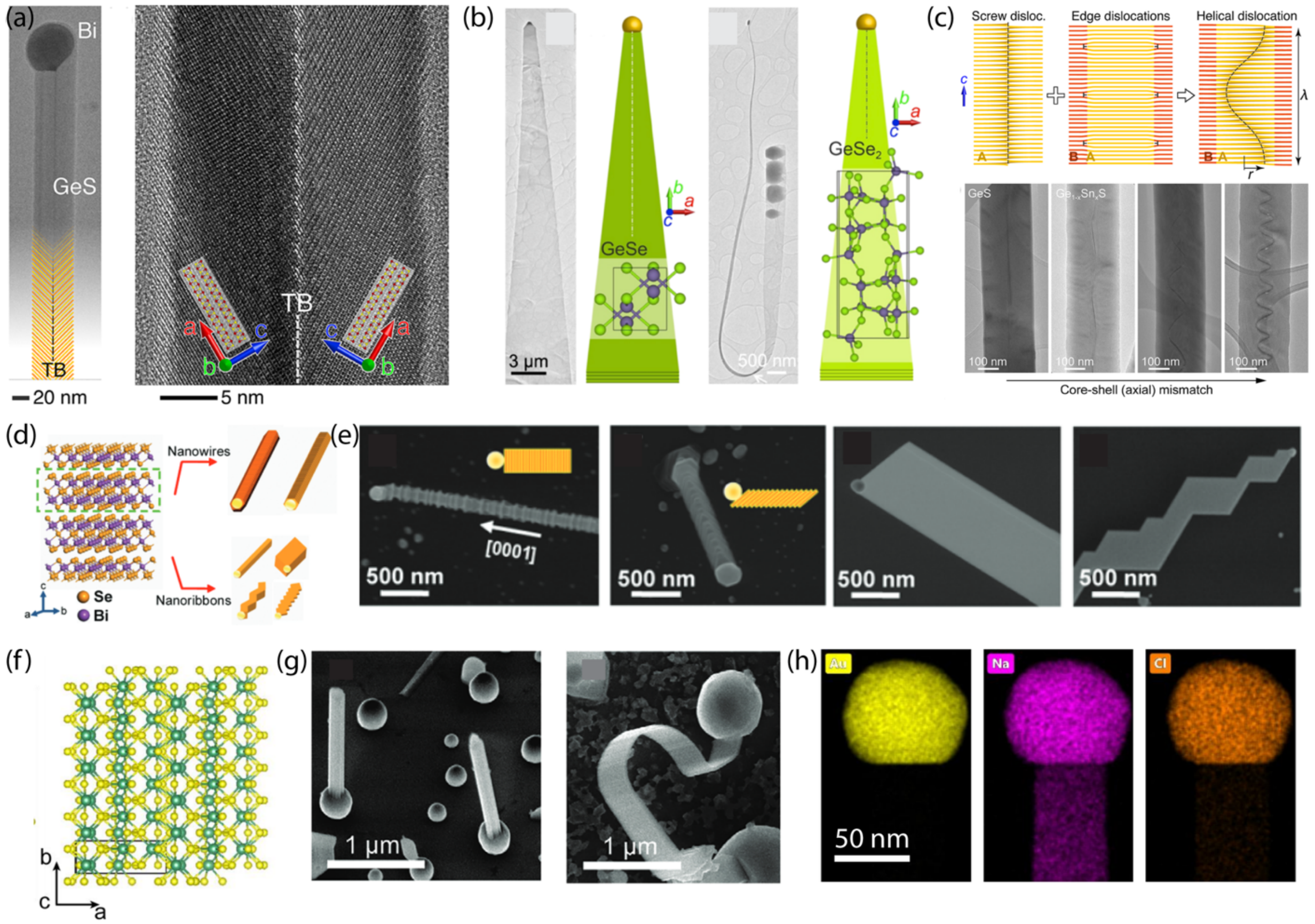

**Figure 9. One-dimensional van der Waals materials by VLS**. (a) Bi-seeded GeS[131] with bicrystal structure and axial screw dislocation. (b) GeSe and $GeSe_2$ nanoribbon with Au seed[132]. (c) Composition-dependent line defects in core-shell $Ge_{1-x}Sn_xS$ heterostructures[113]. (d-e) Nanowires and nanoribbons of Au-$Bi_2Se_3$[30]. (f-h) $NbS_3$ by salt-assisted VLS growth[58]. The seed particle is an alloy of Au, Na and Cl. (a) Adapted with permission from ref[131]. Copyright 2022 American Chemical Society. (b) Adapted with permission from ref[132]. Copyright 2022 American

Chemical Society. (c) Adapted with permission from ref[113]. Copyright 2023 American Chemical Society. (d) Adapted with permission from ref[30]. Copyright 2010 American Chemical Society. (f-h) Adapted with permission from ref[58]. Available under a CC-BY 4.0 license. Copyright The Authors.

The seed particle also selects among competing morphologies and phases. In GeS, the common Au seed produces chiral nanowires [56,111] with axial screw dislocation as mentioned in section 3.2.2 (Fig. 8d). On the other hand, Bi seed changes the dominant fast-growing structures, producing bicrystal GeS ribbons (Fig. 9a) in which a central twin plane accelerates growth[129]. In the Ge-Se system, where stable mono- and dichalcogenide phases coexist, growth conditions select between GeSe and $GeSe_2$ ribbons (Fig. 9b), and post-growth selenization converts ribbon surfaces to form GeSe-$GeSe_2$ core-shell heterostructures[132]. Alloying across the chalcogen or metal site is also accessible, with $GeS_{1-x}Se_x$ (Fig. 8d) or $Ge_{1-x}Sn_xS$ (Fig. 9c), growing from mixed vapors[133,134]. These observations echo the theme of Section 2: the seed is not a passive collector but the component that decides which of several competing structures emerges.

Heterostructure formation deserves specific mention because Section 3.2.3 identified it as largely absent in non-conventional systems. In the monochalcogenides, differences in how fast material incorporates through the droplet and through direct vapor-solid attachment have been exploited to grow mixed-dimensional heterostructures, in which a VLS-grown one-dimensional wire and vapor-solid-grown two-dimensional plates of a different composition form in a single step[133]. This route does not yet provide the abrupt axial junctions available in Si-Ge or GaAs-InAs, but it demonstrates that the coexistence of seeded and non-seeded pathways, treated throughout this review as a control problem, can be turned into a design tool when the two pathways are balanced deliberately.

The layered chalcogenides with thicker van der Waals blocks, $Bi_2Se_3$, $Bi_2Te_3$, and $In_2Se_3$, follow the same logic with a stronger bias toward lateral extension. Au-seeded growth of these materials produces wires, ribbons, and plates, frequently within the same growth run[25,28–30,57], and the seed particles found on ribbons are often undersized or displaced from the ribbon midline (Figs. 9d-e), as discussed in Section 3.2.1. The droplet initiates growth but does not retain exclusive control over it, and vapor-solid incorporation at the layer edges widens the structure as it lengthens. Managing this seeded-to-non-seeded handoff, rather than precursor chemistry or droplet formation, is the central open problem for one-dimensional growth in the layered families.

### 4.3 Transition metal trichalcogenides

The trichalcogenides invert the situation of Section 4.2. Their chain-like bonding supplies the one-dimensional habit that the layered materials must be forced into, and bulk crystals of $NbS_3$, $NbSe_3$, $TaS_3$, $TiS_3$, and $ZrS_3$ have long been grown as needles and fibers by chemical vapor transport[130]. The obstacle is everything upstream of morphology. The constituent metals melt above 2000°C and supply negligible vapor under CVD conditions, so none of the physical routes of Section 1.2 delivers a usable metal flux without assistance. No suitable molecular precursors exist, for the reasons given in Section 1.3. And no binary eutectic between these metals and the common seed materials is accessible in a realistic growth window, so the classical basis for seed selection described in Section 2.1 offers no candidate at all (Table 2).

Salt-assisted VLS growth addresses these failures together rather than one at a time[58] (Figs. 9g-h). Reaction of an alkali halide with the transition metal oxide generates a volatile metal-containing intermediate, as described in Section 1.2, providing the vapor flux that direct evaporation cannot. The same salt participates in the seed. Alkali and halide species are incorporated into the Au particle together with the metal and chalcogen, forming a multicomponent alloyed droplet whose effective melting point and solubility limits differ from those of any binary Au alloy. This second role is the decisive one. An enhanced Nb flux delivered to a solid Au particle would offer no liquid pathway for VLS growth, because no Nb-Au eutectic is accessible between 650 and 800°C. Growth becomes possible because the droplet itself is a different liquid. The distinction drawn in Section 2.2 between salt-assisted VLS and salt-assisted vapor-solid transport, and the precedent of alkali-seeded growth of $MoO_3$ and $MoS_2$ ribbons discussed there, both belong to this same picture of coupled precursor activation and seed chemistry.

The signatures of seeded growth in this system are those established for conventional nanowires in Section 3. Structures terminate in particles whose size correlates with the lateral dimension of the wire or ribbon, small droplets yield narrow wires and larger droplets yield ribbons, and growth location can be set by patterning the seed metal[58]. Within the same growth, wire-like and ribbon-like structures form with distinct crystallographic growth orientations, consistent with two coexisting pathways in which either the droplet or surface diffusion limits the growth rate. The approach extends to $NbSe_3$ and $TiS_3$[58], indicating that the multicomponent droplet strategy is not specific to one compound but applies across the family that shares the same failure of binary seed selection.

The trichalcogenides therefore provide the clearest demonstration of the bottleneck argued in Section 2.5. Precursor activation alone had been available for decades in the form of transport agents and halide chemistry, and the intrinsic anisotropy of the crystals guaranteed one-dimensional habits without any seed. What kept this family outside the reach of controlled nanoscale synthesis was the absence of a workable seed, and what brought it within reach was a change in seed chemistry. At the same time, the multicomponent droplet that enables growth is itself poorly understood. Its phase behavior corresponds to no published phase diagram, its composition during growth is inferred from post-growth analysis, and the questions listed at the end of Section 2.5 apply to it in full.

### 4.4 Twisted and defect-mediated structures in van der Waals nanowires

The general mechanics of screw-dislocation-driven growth and the Eshelby twist were discussed in Section 3.2.2. Van der Waals nanowires give that discussion a specific and unusually productive setting, because an axial screw dislocation in a wire whose layers lie perpendicular to the growth axis converts the stacking sequence into a continuous helix (Figs. 9a and c). The result is an interlayer twist distributed along the wire, and in GeS this twist can be discretized by concurrent radial growth into helical assemblies of nanoplates[56]. Twist angle and twist rate are adjustable through the wire radius, following the Eshelby relation, and through growth chemistry[111,112]. Van der Waals wires thus offer a synthesis-based route to twisted interfaces of the kind otherwise assembled layer by layer from exfoliated flakes, and twist homojunctions with continuously varying interlayer angle have been demonstrated within single wires[120].

Control over the defect itself has advanced beyond the initial demonstrations. Dislocations with mixed screw and edge character have been placed and tuned in core-shell van der Waals wires[113], and helical GeSe/GeS heterostructures have been grown by edge epitaxy on twisted templates[135]. The consequences for transport are now measurable rather than hypothetical. In twisted GeS wires carrying single screw dislocations, the dislocation stabilizes a distinct structural polytype and suppresses room-temperature thermal conductivity by most of its untwisted value at large diameters, while thin twisted wires show a diameter dependence opposite to that of straight wires[114]. These results replace the theoretical expectations cited in Section 3.3 with experimental evidence that dislocation engineering in van der Waals wires modifies functional properties at useful magnitudes.

For the purposes of this review, the relevant point is where these defects originate. As discussed in Section 3.2.2, the nucleation site of the axial dislocation remains unresolved, with hypotheses placing it at the seed particle or at the substrate. In the van der Waals systems the seed is directly implicated in defect selection, since the choice of seed metal shifts the balance between dislocated chiral wires and twin-plane bicrystals in the same material[131]. A seed particle that selects not only phase and morphology but also the defect content of the wire is an opportunity as much as a complication, and it extends the list of seed-controlled outcomes that Section 2 developed for conventional nanowires.

### 4.5 Challenges and opportunities: extending seeded growth to emerging quantum and functional materials

The families discussed above represent only part of a growing set of materials for which one-dimensional form is of interest but seeded growth has not been established (Fig. 10 and Table 1). The first consists of quasi-one-dimensional van der Waals crystals beyond the transition metal trichalcogenides, including ternary transition metal chalcogenides of the form $M_2X_3Y_8$ (M = Ta, Nb; X = Ni, Pd, Pt; Y = S, Se)[136] and III-VI-VII compounds[137] such as GaSeI and InSeI (Figs. 10b-c), together with $Bi_4I_4$[138] and $Sb_2S_3$[139] which host charge density waves, strongly anisotropic transport[136,138,140–142] and intriguing structural twisting behavior. The second group comprises nanowires of quantum materials whose interest lies in correlated and topological electronic phenomena in combination with structural anisotropy, including topological crystalline insulators such as SnTe and SnSe[143], Dirac and Weyl semimetals[144] such as $Cd_3As_2$, TaAs, TaP and WP[145,146] (Fig. 10d), in which superconductivity, charge density waves, magnetism, and nontrivial band topology coexist[147,148]. Last but not least, there is an emerging group of chalcogenide perovskite nanowires $ABX_3$ (A = Ba, Sr; B = Zr, Ti, X = Se, S)[149,150] (Figs. 10e-f) which have showed to exhibit large dielectric constants, high absorption coefficients, and long charge carrier lifetimes, all of which indicate comparable or better properties to the halide perovskites for optoelectronic applications. Most of these materials have been synthesized primarily as bulk crystals by chemical vapor transport or flux methods. Bulk crystals serve fundamental studies well, but they limit the exploration of dimensional confinement effects[29,30,151] and constrain integration with other functional materials. Nanowire realizations would offer control over nanoscale dimension, crystal phase, and doping profile[145], and might open regimes of quantum transport and energy management not accessible in bulk[152]. To date, nearly all nanostructures of these materials that have been measured were obtained without a seed particle, by exfoliation, by chemical vapor transport, or by uncatalyzed vapor-phase growth.

What non-seeded growth does not supply are the controls that organized Sections 1 through 3. Nucleation occurs wherever local conditions favor it, so growth location is not defined. Lateral dimensions are set by kinetics rather than by a droplet footprint, so thickness and width distributions are broad. No mechanism exists for switching the precursor supply during growth, so the axial heterostructures discussed in Section 3.2.3 are out of reach, and with them the compositional modulation and doping control that quantum materials in particular require. Each missing capability corresponds to a function that the seed particle performs in VLS growth of group IV and III-V nanowires.

Whether these capabilities can be transferred to the emerging materials is suggested by a system in which seeded and non-seeded pathways operate side by side. In $Bi_4I_4$, Au particles on the growth substrate act primarily as nucleation sites for vapor-solid growth of nanowires. Particles that instead take up Bi and I in equal proportion during growth produce quasi-two-dimensional sheets of laterally ordered chains through a VLS process[138]. The same material on the same substrate thus yields two morphologies with different growth orientations, and the outcome is decided by the chemical state of the seed particle. Neither the crystal structure nor any binary phase diagram involving Au anticipates the composition at which this transition occurs.

This observation leads back to the bottleneck identified in Section 2.5 (also Table 2). The obstacle to extending seeded growth across these materials is not the materials themselves, and in most cases it is not precursor volatility, since non-seeded routes already crystallize many of them from the vapor. What is missing is a predictive basis for choosing a seed and for anticipating the composition it will adopt under growth conditions. The multicomponent droplets that enabled transition metal trichalcogenide growth[58], which relaxed eutectic constraints for high-melting transition metals and in principle allow the metal precursor to be switched during growth, and the composition-dependent pathway selection in $Bi_4I_4$ [138], point to the same conclusion. Progress will come from determining what the seed particle is during growth, and from developing seed chemistries for the material classes that classical binary selection excludes, not from further expansion of the precursor toolbox.

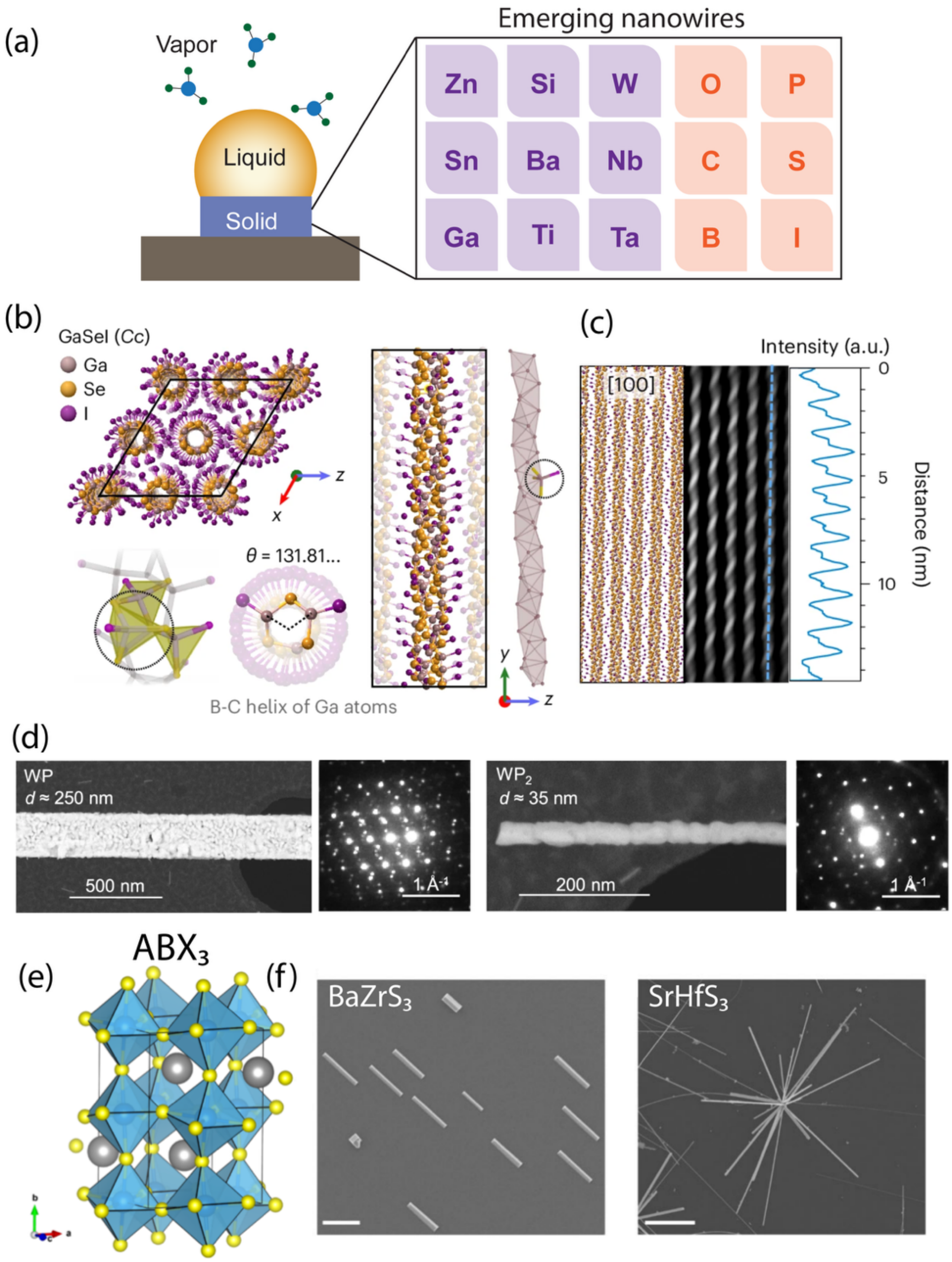

**Figure 10.** (a) Emerging nanowires by VLS synthesis. (b-c) Helical III-VI-VII nanowires exemplified by GaSeI[137]. (d) Topological metals WP and $WP_2$ nanowires[146]. (e) Sulfide perovskite nanowires[150]. (b-c) Adapted with permission from ref [137]. Copyright 2024 Nature Portfolio. (d) Adapted with permission from ref[146]. Available under a Creative Commons CC-BY 4.0 license. Copyright The Authors. (e) Adapted with permission from ref[150]. Copyright 2025 American Chemical Society.

## Summary and Outlook

In this review, we have examined vapor-liquid-solid growth of non-conventional nanowires by following the sequence of the growth process itself: the creation and delivery of precursors, the selection and evolution of seed particles, and the development of one-dimensional structures. At each step we compared oxides, carbides, and chalcogenides against the group IV and III-V baseline, and in Section 4 we applied the full framework to van der Waals materials, whose intrinsic anisotropy places them in a different relationship to the seed particle than any conventional system.

Across these comparisons, the limitations at each step are not equivalent, and we close by stating explicitly which one we consider rate-limiting (Table 2). Precursor creation and delivery constrain which materials can be attempted, and the reactor limitations identified in Section 1.3 are real. But these constraints are increasingly relaxed by activation chemistry, of which the salt-assisted route is one example, and by reactor engineering with independently controlled sources. The corresponding gaps are, in our assessment, tractable engineering problems. The seed particle is different in kind. For conventional nanowires, the binary phase diagram supplied a predictive rule for seed selection, and essentially all deterministic control demonstrated in Si, Ge, and GaAs traces back to knowing and manipulating the state of the droplet. For non-conventional nanowires, no equivalent rule exists. The physical state of the seed during oxide growth remains contested in even the most studied systems, as the conflicting VLS and VSS assignments for Au-seeded ZnO illustrate. The compositions of the multicomponent droplets that enable trichalcogenide growth correspond to no measured phase diagram, and their properties during growth are inferred from post-growth analysis. The seed determines what emerges from the growth, including crystal phase, cross section, defect content, and whether an intended axial heterostructure forms at all, yet it is the component about which the least is known. We identify this, the absence of a predictive and mechanistic understanding of seed particles beyond binary eutectics, as the principal bottleneck restricting deterministic synthesis of non-conventional nanowires.

This conclusion organizes the outlook for the two groups of materials this review has covered. For the oxides, carbides, and chalcogenides that VLS growth has already reached, the accessible morphologies are established but control and understanding lag decades behind the conventional systems. Progress here requires resolving the open questions of Section 2.5, namely the state, composition, and dynamics of the seed during growth. *In situ* microscopy carried this understanding for Si and GaAs, and extending it to these materials depends on growth chemistries and instrumentation that can operate together, a requirement that multicomponent droplets with reduced effective melting temperatures may help satisfy. For the emerging one-dimensional quantum materials of Section 4.5, the situation is complementary. Non-seeded routes already produce high-quality crystals, and what is missing is precisely what a seed provides: defined growth location, dimensional control through the droplet footprint, and the possibility of switching composition mid-growth to form heterostructures. In both directions, the enabling advance is the same, a seed selection framework that extends beyond binary eutectics to alloyed and multicomponent droplets.

Such a framework will not come from intuition alone. Multicomponent droplet compositions span a parameter space too large for exhaustive exploration, which is where the semi-automated growth platforms, searchable synthesis databases, and data-driven analysis discussed in Section 3.3 become necessary rather than convenient. Combined with *in situ* observation of droplet state and composition, these tools could do for multicomponent seeds what the binary phase diagram did for Au-Si: convert seed selection from empirical search into design. The non-conventional nanowires discussed here have resisted deterministic synthesis precisely because that conversion has not yet happened. The materials are identified, their properties are predicted, and in many cases their crystals already grow. What remains is to understand the particle at the tip.

**Acknowledgement:** The authors acknowledge fruitful discussions with Prof. Frances M. Ross (MIT) and Prof. Michael A. Filler (Georgia Tech) in the preparation of this review article.

**Conflict of Interest:** The authors declare no conflict of interest.

TOC Graphic

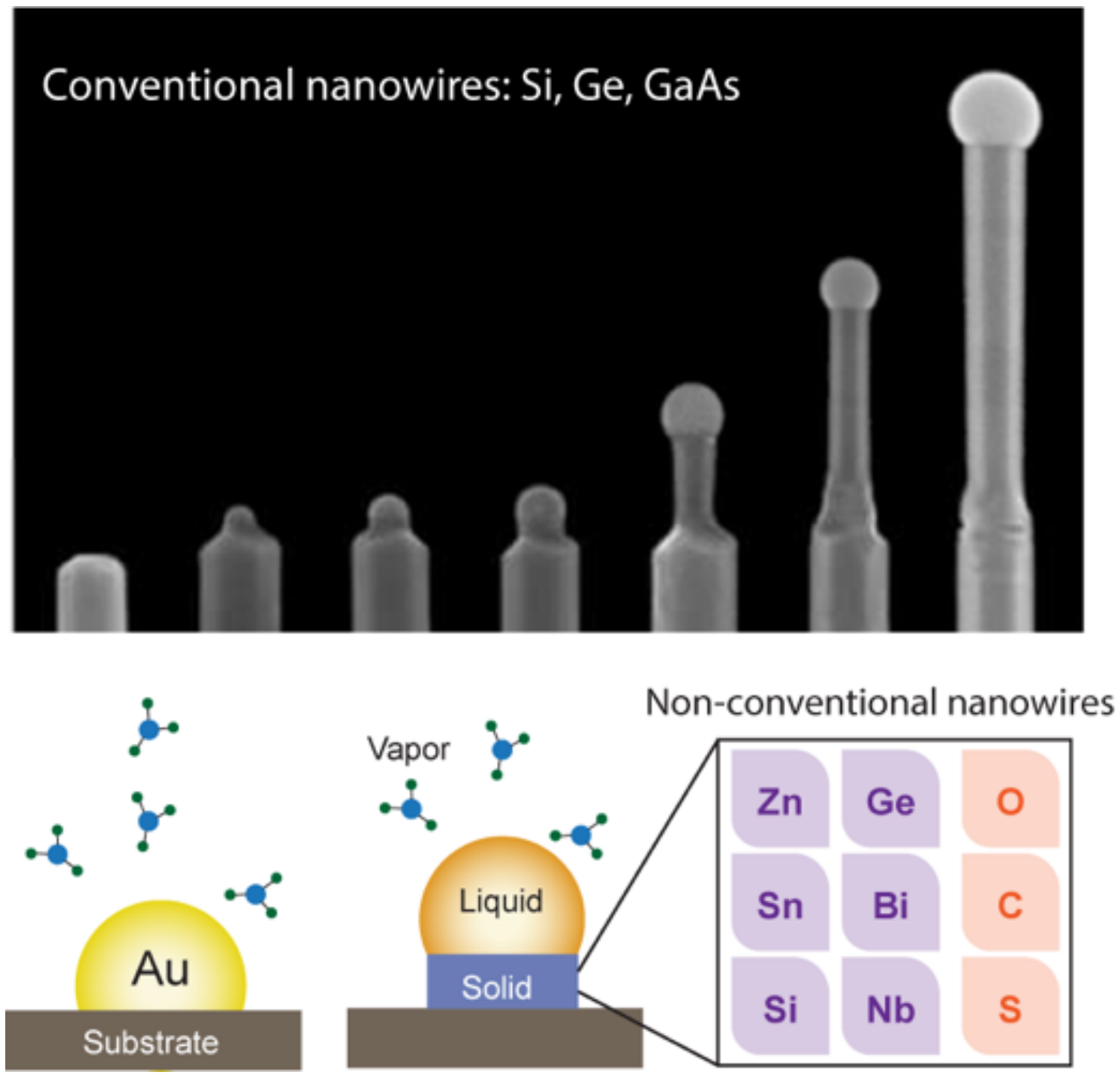
Conventional nanowires: Si, Ge, GaAs
Vapor
Non-conventional nanowires
Au
Substrate
Liquid
Solid
Zn
Ge
O
Sn
Bi
C
Si
Nb
S